%% file: main.tex
\documentclass[accepted]{uai2021} 

\usepackage[american]{babel}

\usepackage{natbib}
\usepackage{mathtools}
\usepackage{booktabs}
\usepackage{tikz}

\usepackage{color}
\usepackage{caption, subcaption}
\usepackage{booktabs}

\usepackage{wrapfig, amsmath, amsthm, mathtools}
\usepackage{bbm}
\usepackage{hyperref}

\usepackage{amsfonts}
\usepackage{amssymb}

\usepackage{bbm}
\usepackage{xcolor}

\usepackage{booktabs}

\usepackage{makecell}
\usepackage{caption, subcaption}

\usepackage{tikz}
\usetikzlibrary{automata,decorations.markings,positioning,arrows}

\DeclareMathOperator{\Tr}{Tr}

\newcommand{\E}{\mathbb{E}}
\newcommand{\Var}{\text{Var}}
\newcommand{\AVar}{\text{AVar}}
\newcommand{\Cov}{\text{Cov}}

\newcommand{\ind}{\perp\!\!\!\!\perp}

\DeclareMathOperator*{\argmin}{arg\,min}

\title{Estimating Treatment Effects with Observed \\ Confounders and Mediators}

\author[ ]{Shantanu~Gupta}
\author[ ]{Zachary~C.~Lipton}
\author[ ]{David~Childers}
\affil[ ]{%
    Carnegie Mellon University \protect \\
    \small{\texttt{\{\href{mailto:shantang@cmu.edu}{shantang},\href{mailto:zlipton@cmu.edu}{zlipton},\href{mailto:dchilders@cmu.edu}{dchilders}\}@cmu.edu}}
}

\begin{document}
\maketitle

\begin{abstract}
\input{sections/00-abstract}
\end{abstract}

\section{Introduction}
\label{sec:introduction}
\input{sections/10-intro.tex}

\section{Related Work}
\label{sec:related}
\input{sections/20-related.tex}

\section{Preliminaries}
\label{sec:preliminaries}
\input{sections/30-preliminaries.tex}

\section{Variance of Backdoor \& Frontdoor Estimators}
\label{sec:frontdoor-backdoor}
\input{sections/40-frontdoor-backdoor.tex}

\section{Combining Mediators \& Confounders}
\label{sec:combined-estimator}
\input{sections/50-combined-estimator.tex}

\section{Combining Revealed-confounder and Revealed-mediator Datasets}
\label{sec:partially-observed}
\input{sections/60-partially-observed.tex}

\section{Experiments}
\label{sec:experiments}
\input{sections/70-experiments.tex}

\section{Discussion}
\label{sec:conclusion}
\input{sections/80-conclusion.tex}

\bibliography{refs}

\appendix
\input{sections/90-appendix.tex}

\end{document}

%% file: sections/00-abstract.tex
Given a causal graph, the do-calculus
can express treatment effects as functionals 
of the observational joint distribution 
that can be estimated empirically.
Sometimes the do-calculus identifies multiple valid formulae,
prompting us to compare the statistical properties 
of the corresponding estimators.
For example, the backdoor formula applies  
when all confounders are observed and
the frontdoor formula applies 
when an observed mediator transmits the causal effect.
In this paper, we investigate the over-identified scenario
where both confounders and mediators are observed,
rendering both estimators valid.
Addressing the linear Gaussian causal model,
we demonstrate that either estimator 
can dominate the other by an unbounded constant factor.
Next, we derive an optimal estimator,
which leverages all observed variables,
and bound its finite-sample variance.
We
show that it
strictly outperforms the backdoor and frontdoor estimators
and that this improvement can be unbounded.
We also present a procedure for combining two datasets,
one with observed confounders and another with observed mediators.
Finally, we evaluate our methods on both 
simulated data and the IHDP and JTPA datasets.

%% file: sections/10-intro.tex
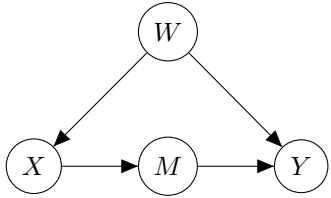
\begin{figure}
\center
  \begin{tikzpicture}[
midarr/.style 2 args={
  decoration={
    markings,
    mark=at position 1 with {\arrow[xshift=0pt]{triangle 45} \node[#1] {#2};},
  },
    postaction=decorate,
},
state/.append style={minimum size=20pt},
node distance=1cm and 1cm
]
\node[state]
  (X) {$X$};
\node[state, right=of X]
  (M) {$M$};
\node[state,right=of M]
  (Y) {$Y$};
\node[state,above=of M]
  (W) {$W$};

\draw[midarr={right=2pt}{}]
  (X) -- (M);  
\draw[midarr={right=2pt}{}]
  (M) -- (Y);  
\draw[midarr={right=2pt}{}]
  (W) -- (X);  
\draw[midarr={below=2pt}{}]
  (W) -- (Y);  
\end{tikzpicture}
{
  \caption{Causal graph with observed mediator and confounder. 
  The backdoor and frontdoor estimators are both applicable.}
  \label{fig:causal_diagram}
}
\end{figure}

Causal effects are not, in general, identifiable from observational data alone.
The fundamental insight of causal inference is that
given structural assumptions on the data generating process,
causal effects may become expressible as functionals
of the joint distribution over observed variables. 
The do-calculus, introduced by \citet{pearl1995causal},
provides a set of three rules that can be used
to convert causal quantities into such functionals.
We are motivated by the observation that,
for some causal graphs, treatment effects may be over-identified.
Here, applications of the do-calculus produce distinct functionals, 
all of which, subject to positivity conditions,
yield consistent estimators of the same causal effect. 
Consider a causal graph (see Figure \ref{fig:causal_diagram})
for which the treatment $X$, mediator $M$, confounder $W$, 
and outcome $Y$ are all observable. 
Using the backdoor adjustment, we can express the 
average treatment effect of $X$ on $Y$ as a function of $P(X,W,Y)$,
while the frontdoor adjustment expresses that 
same causal quantity via $P(X,M,Y)$ \citep{pearl1995causal}.
In our experiments, we work with a real-world dataset that contains both confounders and mediators.
Faced with the (fortunate) condition of overidentification,
our focus shifts from identification: \emph{is our effect estimable?}, to optimality: \emph{which among multiple valid estimators dominates from a standpoint of statistical efficiency?}

In this paper, we address this very graph,
focusing our analysis on the linear causal model \citep{wright1934},
a central object of study in causal inference and econometrics
and also explore the semiparametric setting.
Over-identification can arise in many other causal graphs (e.g. multiple backdoor adjustment sets, multiple instrumental variables, etc.).
However, we focus on this graph because 
the frontdoor estimator is a canonical example 
of a novel identification result derived using graphical models. 
It is central in the causality literature \citep{pearl2018book, imbens2019potential} 
and is a natural first step in the study of over-identified causal models.
Deriving the finite sample variance of the backdoor and frontdoor estimators,
and precisely characterizing conditions under which each dominates,
we find that either may outperform the other to an arbitrary degree
depending on the underlying model parameters.
These expressions can provide guidance to practitioners
for assessing the suitability of each estimator.
For example, one byproduct of our analysis is to characterize
what properties make for the ``ideal mediator''.
Moreover, in the data collection phase, if one has a choice
between collecting data on the mediator or the confounder,
these expressions, together with the practitioner's beliefs
about likely ranges for model parameters,
can be used to decide what data to collect.

Next, we propose techniques that leverage 
both observed confounders and mediators.
For the setting where we \emph{simultaneously} observe
both the confounder and the mediator, 
we introduce an estimator 
that optimally combines all information.
We prove theoretically that this method achieves lower mean squared error (MSE) 
than both the backdoor and frontdoor estimators,
for all settings of the underlying model parameters.
Moreover, the extent to which this estimator can dominate the better 
of the backdoor and frontdoor estimators is unbounded.
Subsequently, we consider the partially-observed setting
in which two datasets are available,
one with observed confounders (but not mediators) $\{(X, W, Y)\}_{i=1}^n$, 
and another with observed mediators (but not confounders) $\{(X, M, Y)\}_{i=1}^m$. 
Interestingly, the likelihood is convex
given simultaneous observations
but non-convex under partially-observed data. 
We introduce
an estimator 
that is guaranteed to achieve higher likelihood
than either the backdoor or frontdoor estimators.
Finally,
we evaluate our methods on  
synthetic, semi-synthetic, and real datasets.
Our proposed estimators that combine confounders and mediators always exhibit lower MSE than the backdoor and frontdoor estimators
when our model assumptions are satisfied.

Our principal contributions are the following:
\begin{enumerate}
\item Derivation of the parameter regimes 
where either of the frontdoor and backdoor estimators dominate
vis-a-vis sample efficiency.
\item Demonstration
of strict (and unbounded) improvements
of the optimal (combined) estimator  
over \emph{both} the frontdoor and backdoor estimators.
\item Adaptation of a semi-parametric estimator to our graph,
showing the benefits of our approach in non-linear settings.
\item Analysis for the partially observed case,
where mediators and confounders are observed separately (but never simultaneously).
\end{enumerate}

%% file: sections/20-related.tex
The backdoor adjustment formalizes the practice 
of controlling for known confounders
and is widely applied in statistics and econometrics \citep{pearl2009causality, pearl2010foundations,perkovic2015complete}.
The frontdoor adjustment,
which leverages observed mediators to identify causal effects
even amid unobserved confounding, 
has seen increasing application in real-world datasets
\citep{bellemare2019paper, glynn2018front, glynn2017front, chinco2016misinformed,cohen2014friends}. 

In the most similar work to ours, 
\citet{glynn2018front} compare the frontdoor and backdoor adjustments,
computing bias (but not variance) formulas for each 
and performing sensitivity analysis.
Exploring a real-world job training dataset,
they demonstrate that the frontdoor estimator 
outperforms its backdoor counterpart (in terms of bias).
The finite sample variance of the frontdoor estimator for the linear Gaussian case was previously derived by \citet{kuroki2000selection}.
\citet{ramsahai2012supplementary} compare the frontdoor and backdoor estimators based on their asymptotic variances and also show that the combined estimator's variance cannot be higher than the other two.
\citet{kuipers2020variance} derive the finite-sample variances of two possible adjustments in a three-variable binary causal graph
and show that the optimal estimator depends on the model parameters.
\citet{henckel2019graphical} introduce a graphical criterion for comparing the asymptotic variances of adjustment sets for the backdoor criterion in linear causal models.
\citet{rotnitzky2019efficient} extend this work,
showing that the same graphical criterion is valid 
for non-parametric causal models.
They also present a semi-parametric efficient estimator 
that exploits the conditional independencies in a causal graph.

Researchers have also worked 
to generalize the frontdoor criterion.
\citet{bareinboim2019causal} introduce 
the conditional frontdoor criterion,
allowing for both treatment-mediator confounders
and mediator-outcome confounders. 
\citet{fulcher2020robust} propose a method 
for including observed confounders along with a mediator 
with discrete treatments.

The study of overidentified models 
dates at least back to  \citet{koopmans1950identification}. 
\citet{sargan1958estimation, Hansen1982} formalized the result 
that in the presence of overidentification, 
multiple estimators can be combined to improve efficiency. 
This was extended to the non-parametric setting by \citet{chen2018overidentification}.
A related line of work considers methods 
for combining multiple datasets for causal inference. 
\citet{bareinboim2016causal} study the problem 
of handling biases while combining heterogeneous datasets, 
while \citet{jackson2009bayesian} present Bayesian methods 
for combining datasets with different covariates and some common covariates.

%% file: sections/30-preliminaries.tex
In this work, we work within the structural causal model (SCM) framework due to  \citet{pearl2009causality},
formalizing causal relationships via directed acyclic graphs (DAGs).
Each $X \rightarrow Y$ edge in this DAG indicates that the variable $X$ 
is (potentially) a direct cause of variable $Y$.
All measured variables are deterministic functions of their parents and a set of jointly independent per-variable noise terms. 

\noindent \textbf{Linear Gaussian SCM \quad} In linear Gaussian SCMs,
each variable is assumed to be a linear function of its parents. 
The noise terms are assumed to be additive and Gaussian. 
In this paper,
the finite sample results are derived for
the linear Gaussian SCM 
for the overidentified confounder-mediator graph 
(Figure \ref{fig:causal_diagram}),
where the structural equations can be written as
\begin{align} \label{eq:linearSCM}
\begin{split}
    w_i &= u^w_i, \\
    x_i &= d w_i + u^x_i, \\
    m_i &= c x_i + u^m_i, \\
    y_i &= a m_i + b w_i + u^y_i,
\end{split}
\begin{split}
    u^w_i &\sim \mathcal{N}(0, \sigma^2_{u_w}) \\
    u^x_i &\sim \mathcal{N}(0, \sigma^2_{u_x}) \\
    u^m_i &\sim \mathcal{N}(0, \sigma^2_{u_m}) \\
    u^y_i &\sim \mathcal{N}(0, \sigma^2_{u_y}).
\end{split}
\end{align}
Here, $w_i, x_i, m_i$, and $y_i$ are realized values 
of the random variables $W, X, M, Y$, respectively,
and $u^w_i, u^x_i, u^m_i, u^y_i$ are realized values 
of the corresponding noise terms.
The zero mean assumption in Eq. \ref{eq:linearSCM} simplifies analysis, but is not necessary for the results presented in this paper.

\subsection{The Backdoor and Frontdoor Adjustments}

The effect of a treatment $X$ is expressible 
in terms of the post-intervention distributions of 
the outcome $Y$ for different values of the treatment $X=x$.
An intervention $do(X=x)$ in a causal graph
can be expressed via the \emph{mutilated graph} that results 
from deleting all incoming arrows to $X$, 
setting $X$'s value to $X=x$ for all instances, 
while keeping the SCM otherwise identical. 
This distribution is denoted as $P(Y|do(X=x))$.

The backdoor and frontdoor adjustments \citep{pearl2009causality} 
express treatment effects
as functionals of the observational distribution. 
Consider our running example of the causal model in Figure \ref{fig:causal_diagram}. 
We denote $X$ as the treatment, $Y$ as the outcome,
$W$ as a confounder, and $M$ as a mediator. 
Our goal is to estimate the causal quantity $P(Y|do(X=x))$.

\noindent \textbf{Backdoor Adjustment  \quad} 
When all confounders of both $X$ and $Y$ are observed---in 
our example, $W$---then 
the causal effect of $X$ on $Y$, i.e., $P(Y|do(X=x))$ can be written as
\begin{align} \label{eq:backdoor_adjustment_probability_notation}
    P(Y|&do(X=x)) \nonumber \\
        &= \sum_{w} P(Y|X=x,W=w)P(W=w).
\end{align}
\noindent \textbf{Frontdoor Adjustment  \quad} 
This technique applies even when the confounder $W$ is unobserved.
Here we require access to a mediator $M$ 
that (i) is observed;
(ii) transmits the entire causal effect from $X$ to $Y$;
and (iii) is not influenced by the confounder $W$ given $X$. 
The effect of $X$ on $Y$ is computed in two stages. 
We first find the effect of $X$ on $M$, then the effect of $M$ on $Y$ as:
\begin{align}
    & P(M=m|do(X=x)) = P(M=m|X=x) \label{eq:frontdoor_adjustment_probability_x_on_m} \\
    & \begin{aligned}
        P(Y|&do(M=m)) \\
            &= \sum_{x} P(Y|M=m,X=x)P(X=x).
    \end{aligned} \label{eq:frontdoor_adjustment_probability_m_on_y}
\end{align}
We can then write the causal effect of $X$ on $Y$ as
\begin{align*}
    P(&Y|do(X=x)) \\
        &= \sum_{m} P(M=m|do(X=x)) P(Y|do(M=m)).
\end{align*}

%% file: sections/40-frontdoor-backdoor.tex
In this section, we analyze 
the backdoor and frontdoor estimators and characterize the regimes where each dominates. 
We work with the linear SCM described in Eq. \ref{eq:linearSCM}. 
Throughout, our goal is to estimate 
the causal effect of $X$ on $Y$.
In terms of the underlying parameters of the linear SCM, 
the quantity that we wish to estimate is $ac$. 
Absent measurement error, both estimators are unbiased (see proof in Appendix \ref{sec:appendix-unbiasedness})
and thus we focus our comparison on their respective variances.

\noindent \textbf{Variance of the Backdoor Estimator \quad}
The backdoor estimator requires only that we observe $\{X, Y, W\}$
(but not necessarily the mediator $M$).
Say we observe the samples $\{x_i, y_i, w_i\}^n_{i=1}$.
We can estimate the causal effect $ac$ 
by taking the coefficient on $X$ in an OLS regression of $Y$ on $\{X, W\}$. 
This controls for the confounder $W$ 
and corresponds naturally to the adjustment described 
in Eq.~\ref{eq:backdoor_adjustment_probability_notation}.

The finite sample and asymptotic variances of the backdoor estimator are (see proof in Appendix \ref{sec:appendix-variance-results-backdoor})
\begin{align}
    & \Var(\widehat{ac})_{\text{backdoor}} = \frac{a^2 \sigma^2_{u_m} + \sigma^2_{u_y}}{(n-3)\sigma^2_{u_x}}, \nonumber \\
    & \lim_{n \to \infty} \Var(\sqrt{n}(\widehat{ac} - ac))_{\text{backdoor}} =
        \frac{a^2 \sigma^2_{u_m} + \sigma^2_{u_y}}{\sigma^2_{u_x}}.
        \label{eq:backdoor_variance}
\end{align}

\noindent \textbf{Variance of the Frontdoor Estimator \quad}
The frontdoor estimator is used when $\{X, Y, M\}$ samples
are observed.
Say we observe the samples $\{x_i, y_i, m_i\}^n_{i=1}$.
First, we estimate $c$ by taking the coefficient on $X$ in an OLS regression of $M$ on $X$. Let the estimate be $\widehat{c}$. This corresponds to the adjustment in Eq. \ref{eq:frontdoor_adjustment_probability_x_on_m}. Then, we estimate $a$ by taking the coefficient on $M$ in an OLS regression of $Y$ on $\{M, X\}$. Let the estimate be $\widehat{a}_f$. This corresponds to the adjustment in Eq. \ref{eq:frontdoor_adjustment_probability_m_on_y}.

The finite sample variances of $\widehat{c}$ and $\widehat{a}_f$ are (see proof in Appendix \ref{sec:appendix-variance-results-frontdoor})
\begin{align}
    & \Var(\widehat{c}) = \frac{\sigma^2_{u_m}}{(n-2)(d^2 \sigma^2_{u_w} + \sigma^2_{u_x})}, \\
    & \Var(\widehat{a}_f) = \frac{b^2 \sigma^2_{u_w} \sigma^2_{u_x} + \sigma^2_{u_y} (d^2 \sigma^2_{u_w} + \sigma^2_{u_x})}{(n-3) (d^2 \sigma^2_{u_w} + \sigma^2_{u_x}) \sigma^2_{u_m}} \label{eq:frontdoor-a-and-c-hat-variance}.
\end{align}
Using the facts that $\text{Cov}(\widehat{a}_f, \widehat{c}) = 0$ and $\Cov(\widehat{a}_f^2, \widehat{c}^2) = \Var(\widehat{a}_f)\Var(\widehat{c})$, the finite sample variance of the frontdoor estimator is (see proof in Appendix \ref{sec:appendix-frontdoor-finite-sample-variance-result})
\begin{align} \label{eq:frontdoor-finite-sample-variance}
    \Var(\widehat{a}_f\widehat{c}) &= c^2 \Var(\widehat{a}_f) + a^2 \Var(\widehat{c}) + 2 \Var(\widehat{a}_f) \Var(\widehat{c}).
\end{align}
And the asymptotic variance, which does not require Gaussianity, is (see proof in Appendix \ref{sec:appendix-frontdoor-asymptotic-variance-result})
\begin{align}\label{eq:frontdoor_asymp_var}
    \lim_{n \to \infty} & \Var(\sqrt{n}(\widehat{a}_f\widehat{c} - ac)) =  \frac{c^2 (b^2 \sigma^2_{u_w} \sigma^2_{u_x} + \sigma^2_{u_y} D)}{D \sigma^2_{u_m}} + \frac{a^2 \sigma^2_{u_m}}{D}, \nonumber \\
    & \text{where}\,\, D = d^2 \sigma^2_{u_w} + \sigma^2_{u_x}.
\end{align}

\noindent \textbf{The Ideal Frontdoor Mediator \quad}
A natural question then arises:
what properties of a mediator 
make the frontdoor estimator most precise?
We can see that $\Var(\widehat{a}_f \widehat{c})$ is non-monotonic in the mediator noise $\sigma_{u_m}$.
Eq. \ref{eq:frontdoor-finite-sample-variance} provides us with guidance.
$\Var(\widehat{a}_f \widehat{c})$ is a convex function of $\sigma^2_{u_m}$.
The \emph{ideal mediator}
will have noise variance $\sigma^{2*}_{u_m}$ which minimizes Eq. \ref{eq:frontdoor-finite-sample-variance}. That is,
\begin{align*}
    \sigma^{2*}_{u_m} &= \argmin_{\sigma^2_{u_m}} \left[ \Var(\widehat{a}_f\widehat{c}) \right] \\
    \implies \sigma^{2*}_{u_m} &= \frac{|c|\sqrt{b^2 \sigma^2_{u_w} \sigma^2_{u_x} + \sigma^2_{u_y} D }}{|a|} \sqrt{\frac{n-2}{n-3}},
\end{align*}
where $D = d^2 \sigma^2_{u_w} + \sigma^2_{u_x}$.

\noindent \textbf{Comparison of Backdoor and Frontdoor Estimators \quad}
The relative performance of the backdoor and frontdoor estimators depend on the underlying SCM's parameters. 
Using Eqs. \ref{eq:backdoor_variance} and \ref{eq:frontdoor-finite-sample-variance}, the ratio of the backdoor to frontdoor variance is
\begin{align}\label{eq:ratio-backdoor-to-frontdoor}
    & R_{\text{Var}} = \frac{\text{Var}(\widehat{ac})_{\text{backdoor}}}{\text{Var}(\widehat{a}_f\widehat{c})} \\
        &= \frac{(n-2)\sigma^2_{u_m} D^2 (a^2 \sigma^2_{u_m} + \sigma^2_{u_y})}{\sigma^2_{u_x}((n-3)a^2\sigma^4_{u_m}D + (2 \sigma^2_{u_m} + c^2 (n-2) D ) E)} \nonumber,
\end{align}
where $D = (d^2 \sigma^2_{u_w} + \sigma^2_{u_x})$ and $E = (b^2 \sigma^2_{u_w} \sigma^2_{u_x} + \sigma^2_{u_y} D)$.
The backdoor estimator dominates 
when $R_{\text{Var}} < 1$
and vice versa when $R_{\text{Var}} > 1$.
Note that there exist parameters
that cause any value of $R_{\text{Var}} > 0$. 
In particular, as $\sigma^2_{u_x} \rightarrow 0$, $R_{\text{Var}} \rightarrow \infty$ and as $\sigma^2_{u_x} \rightarrow \infty$, $R_{\text{Var}} \rightarrow 0$,
regardless of the sample size $n$. 
Thus, either estimator can dominate the other by any arbitrary constant factor.

%% file: sections/50-combined-estimator.tex
Having characterized the performance of each estimator separately, we now consider optimal strategies for estimating treatment effects in the overidentified regime, where we observe both the confounder and the mediator simultaneously.
Say we observe $n$ samples $\{x_i, y_i, w_i, m_i\}_{i=1}^{n}$. 
We show that the maximum likelihood estimator (MLE) 
is strictly better than the backdoor and frontdoor estimators. The MLE will be optimal since our model satisfies 
the necessary regularity conditions for MLE optimality (by virtue of being linear and Gaussian).
The combined estimator is unbiased (see Appendix~\ref{sec:appendix-unbiasedness-combined}) and thus we focus on the variance.

Let the vector $\mathbf{s_i} = [x_i, y_i, w_i, m_i]$ denote the $i^{\text{th}}$ sample.
Since the data is multivariate Gaussian, 
the log-likelihood of the data is $\mathcal{LL} = -\frac{n}{2} \left[\log{(\det{\Sigma})} + \Tr{(\widehat{\Sigma} \Sigma^{-1})}\right]$,
where $\Sigma = \text{Cov}([X, Y, W, M])$
and $\widehat{\Sigma} = \frac{1}{n} \sum_{i=1}^{n} \mathbf{s_i} \mathbf{s_i}^\top$.
The MLE for a Gaussian graphical model is $\Sigma^{\text{MLE}} = \widehat{\Sigma}$ \citep{uhler2019gaussian}. Let the MLE estimates for parameters $c$ and $a$ be $\widehat{c}$ and $\widehat{a}_c$, respectively. Then
\begin{align}
    \widehat{c} &= \frac{\widehat{\Sigma}_{1,4}}{\widehat{\Sigma}_{1,1}}, \,\,\, \widehat{a}_c = \frac{\widehat{\Sigma}_{1,4}\widehat{\Sigma}_{3,3} - \widehat{\Sigma}_{1,3}\widehat{\Sigma}_{3,4}}{\widehat{\Sigma}_{3,3}\widehat{\Sigma}_{4,4} - \widehat{\Sigma}^2_{3,4}} \label{eq:combined-a-and-c-hat-mle}.
\end{align}
The MLE estimate for $c$ in Eq. \ref{eq:combined-a-and-c-hat-mle} 
is the same as for the frontdoor---the
coefficient of $X$ in an OLS regression of $M$ on $X$. 
The MLE estimate for $a$ in Eq. \ref{eq:combined-a-and-c-hat-mle} 
is the coefficient of $M$ in an OLS regression of $Y$ on $\{M, W\}$. 
The finite sample variance of $\widehat{a}_c$ is (see proof in Appendix \ref{sec:appendix-combined-a-hat-variance})
\begin{align}
    \Var(\widehat{a}_c) &= \frac{\sigma^2_{u_y}}{(n-3)(c^2 \sigma^2_{u_x} + \sigma^2_{u_m})} \label{eq:bothdoor-a-hat-variance}.
\end{align}
The variance of $\widehat{c}$ is the same 
as the frontdoor case as in Eq. \ref{eq:frontdoor-a-and-c-hat-variance}.
Let $r_1 = \sqrt{\frac{n-3}{n-5}}$, $r_2 = \sqrt{\frac{3(n-2)}{n-4}}$, and $L = \left( \frac{c^2 \sigma^2_{u_y}}{c^2 \sigma^2_{u_x} + \sigma^2_{u_m}} + \frac{a^2 \sigma^2_{u_m}}{d^2 \sigma^2_{u_w} + \sigma^2_{u_x}} \right)$. We can bound the finite sample variance of the combined estimator as
\begin{align} \label{eq:combined-finite-sample-upper-bound}
    \frac{L}{n} & \leq \Var(\widehat{a}_c \widehat{c}) \leq \left( c^2 \Var(\widehat{a}_c) + a^2 \Var(\widehat{c}) +  r_1 \right. \nonumber \\
    & \left. \left( 2 |c| \Var(\widehat{a}_c) \sqrt{\Var(\widehat{c})} + r_2 \Var(\widehat{a}_c) \Var(\widehat{c}) \right) \right).
\end{align}
The lower bound is derived using the Cramer-Rao theorem (since the estimator is unbiased) and for the upper bound, we use the Cauchy-Schwarz inequality. The complete proof is in Appendix \ref{sec:appendix-combined-variance-finite-sample}.
And the asymptotic variance, which does not require Gaussianity, is (see proof in Appendix \ref{sec:appendix-combined-asymptotic-variance})
\begin{align}\label{eq:bothdoor_asymp_var}
    \lim_{n \to \infty} \Var(\sqrt{n}(\widehat{a}_c\widehat{c} - ac)) &= L.
\end{align}

\noindent \textbf{The Ideal Mediator \quad}
Just as with the frontdoor estimator, 
we can ask what makes for an \emph{ideal mediator} 
in this case. 
Eq. \ref{eq:bothdoor_asymp_var} shows that $\lim_{n \to \infty} \Var(\sqrt{n}\widehat{a}_c\widehat{c})$
is a convex function of $\sigma^2_{u_m}$.
The ideal mediator will have noise variance $\sigma^{2*}_{u_m}$ which minimizes the variance in Eq. \ref{eq:bothdoor_asymp_var}. 
We use the asymptotic variance here since we only have finite-sample bounds on the variance of the combined estimator.
This means that
\begin{align*}
    & \sigma^{2*}_{u_m} = \argmin_{\sigma^2_{u_m}} \left[ \lim_{n \to \infty} \Var(\sqrt{n}\widehat{a}_c\widehat{c}) \right] \\
    \implies & \sigma^{2*}_{u_m} = \max \left\{0, \frac{|c| \sigma_{u_y} \sqrt{d^2 \sigma^2_{u_w} + \sigma^2_{u_x} }}{|a|} - c^2 \sigma^2_{u_x} \right\}.
\end{align*}

\subsection{Comparison with Backdoor and Frontdoor Estimators}
\label{sec:combined-estimator-dominates-better-of-both}
We can compare Eqs. \ref{eq:backdoor_variance} and \ref{eq:bothdoor_asymp_var} too see that, asymptotically, the combined estimator has lower variance than the backdoor estimator for all values of model parameters. That is, as $n \rightarrow \infty$, $\Var(\sqrt{n}\widehat{a}_c \widehat{c}) \leq \Var(\sqrt{n}\widehat{ac})_{\text{backdoor}}$.
Similarly, we can compare Eqs. \ref{eq:frontdoor_asymp_var} and \ref{eq:bothdoor_asymp_var} to see that, asymptotically, the combined estimator is always better than the frontdoor estimator for all values of model parameters. That is, as $n \rightarrow \infty$, $\Var(\sqrt{n}\widehat{a}_c \widehat{c}) \leq \Var(\sqrt{n}\widehat{a}_f \widehat{c})$.

In the finite sample case, using Eqs. \ref{eq:backdoor_variance} and \ref{eq:combined-finite-sample-upper-bound}, we can see
that for all model parameters,
for a large enough $n$,
the combined estimator will dominate the backdoor.
That is, $\exists N, \text{s.t.}, \forall n > N, \Var(\widehat{a}_c \widehat{c}) \leq \Var(\widehat{ac})_{\text{backdoor}}$,
where the dependence of $N$ on the model parameters is stated in Appendix \ref{sec:appendix-combined-better-than-backdoor}. We can make a similar argument for the dominance of the combined estimator over the frontdoor estimator. Using Eqs \ref{eq:frontdoor-finite-sample-variance} and \ref{eq:combined-finite-sample-upper-bound}, it can be shown that $\exists N, \text{s.t.}, \forall n > N, \Var(\widehat{a}_c \widehat{c}) \leq \Var(\widehat{a}_f \widehat{c})$,
where the dependence of $N$ on the model parameters is stated in Appendix \ref{sec:appendix-combined-better-than-frontdoor}.

Next, we show that the combined estimator can dominate the better of the backdoor and frontdoor estimators by an arbitrary amount. That is, we show that the quantity $R = \frac{\min \left\{ \Var(\widehat{ac})_\text{backdoor}, \Var(\widehat{a}_f\widehat{c}) \right\}}{\Var(\widehat{a}_c\widehat{c})}$
is unbounded. Consider the case when $\Var(\widehat{ac})_\text{backdoor} = \Var(\widehat{a}_f\widehat{c})$.
This condition holds for certain settings of the model parameters
(see Appendix \ref{sec:appedix-combined-estimator-dominates-better-of-both} for an example). 
Here,
\begin{align}
    R &= \frac{\Var(\widehat{ac})_\text{backdoor}}{\Var(\widehat{a}_c\widehat{c})} \nonumber \\
    & \geq \frac{(n-2) D E (a^2 \sigma^2_{u_m} + \sigma^2_{u_y})}{ \sigma^2_{u_x} \left( F + \sigma^2_{u_y} \left( \sigma^2_{u_m} + \sqrt{3} \sigma^2_{u_m} H  \right) \right) }, \label{eq:ratio-combined-unbounded-better}
\end{align}
where $D = d^2 \sigma^2_{u_w} + \sigma^2_{u_x}$, $E = c^2 \sigma^2_{u_x} + \sigma^2_{u_m}$, $r_1 = \sqrt{\frac{n-3}{n-5}}$, $r_2 = \sqrt{\frac{n-2}{n-4}}$, $F = (n-3) a^2 \sigma^2_{u_m} E$, $G = r_1 \frac{\sigma_{u_m}}{ \sqrt{(n-2) D} }$, $H = r_1 r_2 + |c| (n-2) D \left( |c| + G \right)$ and, in Eq. \ref{eq:ratio-combined-unbounded-better}, we used Eq. \ref{eq:combined-finite-sample-upper-bound}. We can see that as $\sigma_{u_x} \rightarrow 0$, $R \rightarrow \infty$ and thus $R$ is unbounded. This shows that, even in finite samples, combining confounders and mediators can lead to an arbitrarily better estimator than the better of the backdoor and frontdoor estimators.

\subsection{Semi-Parametric Estimators} \label{sec:influence-function-estimators}

\citet{fulcher2020robust} derive the efficient influence function
and semi-parametric efficiency bound for a generalized model with discrete treatment and non-linear relationships between the variables. While they allow for confounding of the treatment-mediator link and the mediator-outcome link, the graph in Figure \ref{fig:causal_diagram} has additional restrictions.
As per \citet{chen2018overidentification}, this graph is \textit{locally overidentified}.
This suggests that it is possible to improve the estimator by \citet[Eq. (6)]{fulcher2020robust} (which we refer to as IF-Fulcher).
In our model, there are two additional conditional independences compared to the graph studied in \citet{fulcher2020robust}: $Y \ind X | (M, W)$, and $M \ind W | X$.
We incorporate these conditional independences in IF-Fulcher
by using $\E[Y|M,W,X]=\E[Y|M,W]$, and $f(M|X,W)=f(M|X)$ to create an estimator we refer to as
IF-Restricted:
\begin{align*}
    & \widehat{\Psi} = \frac{1}{n} \sum_{i=1}^n (Y_i-\widehat{\mathbb{E}}[Y|M_i,W_i])\frac{\widehat{f}(M|x^*)}{\widehat{f}(M|X_i)} + \nonumber \\
    & \frac{1\{X_i=x^*\}}{\widehat{P}(X_i=x^*|W_i)} \times \left\{\widehat{\mathbb{E}}[Y|M_i,W_i]- \nonumber \right. \\
    & \left. \sum_{m}\mathbb{\widehat{E}}[Y|m,W_i]f(m|X_i)\right\} + \sum_{m}\widehat{\mathbb{E}}[Y|m,W_i]\widehat{f}(m|x^*),
\end{align*}
where, if $\widehat{f}, \widehat{P}$, and $\widehat{\E}$ are consistent estimators, then $\widehat{\Psi} \overset{p}{\to} \mathbb{E}[Y|do(X=x^*)]$.
By double robustness of the given estimator, if $\widehat{f}, \widehat{P}$, and $\widehat{\E}$ are correctly specified, then IF-Restricted has identical asymptotic distribution as IF-Fulcher.
But using the additional restrictions improves estimation of nuisance functions.
Thus we expect the proposed semi-parametric estimator to perform better in finite samples.
\citet{rotnitzky2019efficient}, in contemporaneous work, analyzed the same graph and showed that, in addition, the efficient influence function is also changed when imposing these conditional independences (see Example 10 in their paper) (we refer to the estimator for this influence function as IF-Rotnitzky).
For our experiments with binary treatments, we use linear regression for $\widehat{f},\, \widehat{\mathbb{E}}$ and logistic regression for $\widehat{P}$.
Another way to adapt IF-Fulcher is for the case when we do not observe the confounders (as in the frontdoor adjustment). In this case, we can set $W_i = \varnothing$ and apply $\widehat{\Psi}$. We call this special case IF-Frontdoor.

%% file: sections/60-partially-observed.tex
We now consider a situation in which
the practitioner has access to two datasets.
In the first one, the confounders are observed but the mediators are unobserved. 
In the second one, the mediators are observed but the confounders are unobserved. 
This situation might arise if data is collected by two groups,
the first selecting variables to measure to apply the backdoor adjustment
and the second selecting variables to apply the frontdoor adjustment.
Given the two datasets,
we wish to optimally 
leverage all available data to estimate the effect of $X$ on $Y$.

A naive approach would be to apply the backdoor and frontdoor estimator
to the first and second dataset, respectively, 
and take a weighted average of the two estimates. 
However, in this case, the variance will be between that of 
the frontdoor and backdoor estimator.
We analyze the MLE,
showing
that this estimator 
has lower asymptotic variance 
than both the backdoor and frontdoor estimators.

\noindent \textbf{Combined Log-Likelihood under Partial Observability \quad}
Say we have $P$ samples of $\{x_i, y_i, w_i\}^P_{i=1}$. 
Let each such sample be denoted by the vector $\mathbf{p}_i = [x_i, y_i, w_i]$.
Moreover, say we have $Q$ samples of $\{x_i, y_i, m_i\}^Q_{i=1}$. 
Let each such sample be denoted using the vector $\mathbf{q}_j = [x_j, y_j, m_j]$. 
Let the observed data be represented as $D$. That is, $D = \left\{ \mathbf{p}_1, \mathbf{p}_2, \hdots, \mathbf{p}_P, \mathbf{q}_1, \mathbf{q}_2, \hdots, \mathbf{q}_Q \right\}$.
Let $N = P + Q$ and let $k = \frac{P}{N}$.
Since the data is multivariate Gaussian, the conditional log-likelihood given $k$ can be written as
\begin{align}\label{eq:partial_log_likelihood}
    \mathcal{LL}(D|k) &= -\frac{N}{2}[k  \left( \log{\det{\Sigma_p}} + \Tr{(\widehat{\Sigma}_p \Sigma^{-1}_p)} \right) + \nonumber \\
    & (1-k) \left( \log{\det{\Sigma_q}} + \Tr{(\widehat{\Sigma}_q \Sigma^{-1}_q)} \right) ],
\end{align}
where $\Sigma_p = \text{Cov}([X, Y, W])$, $\Sigma_q = \text{Cov}([X, Y, M])$, 
$\widehat{\Sigma}_p = \frac{\sum_{i=1}^P \mathbf{p}_i \mathbf{p}_i^\top}{P} $ and $\widehat{\Sigma}_q = \frac{\sum_{i=1}^Q \mathbf{q}_i \mathbf{q}_i^\top}{Q}$.

\begin{figure}[t]
\begin{center}
\begin{subfigure}{0.48\columnwidth}
\includegraphics[width=0.99\columnwidth]{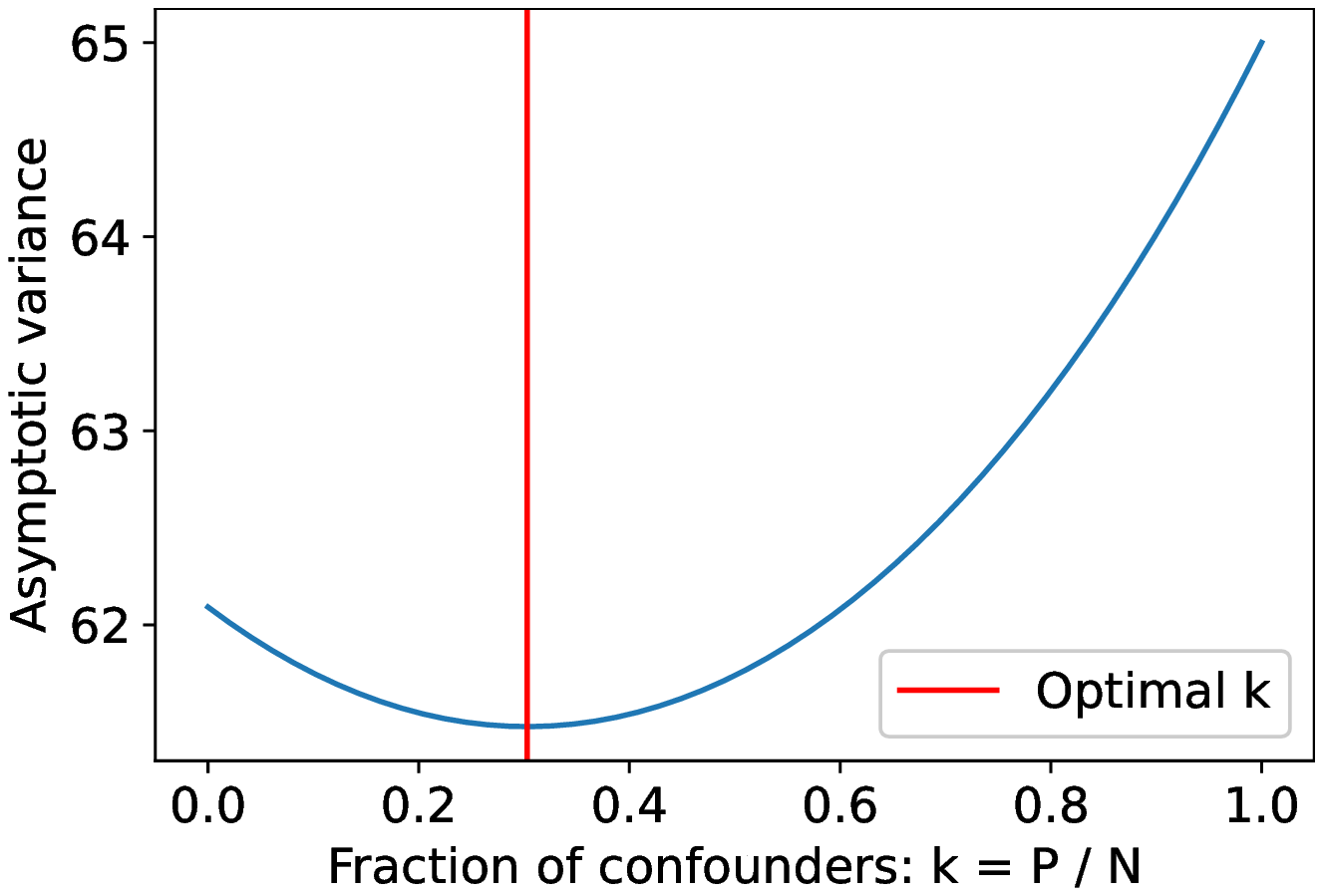}
\caption{Optimal k = $0.3$}
\label{fig:partial-cramer-rao-a}
\end{subfigure}
\begin{subfigure}{0.48\columnwidth}
\includegraphics[width=0.99\columnwidth]{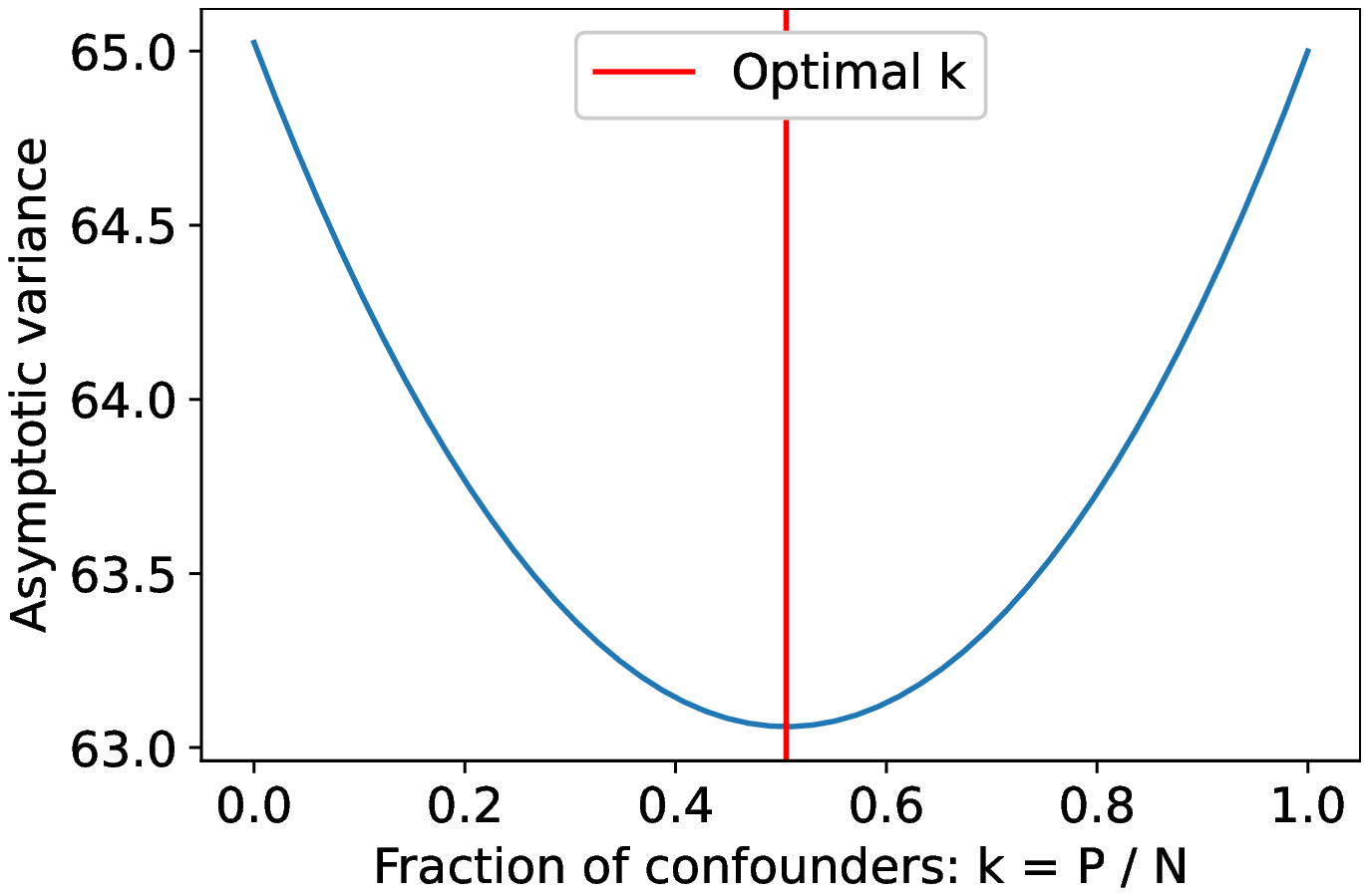}
\caption{Optimal k = $0.5$}
\label{fig:partial-cramer-rao-b}
\end{subfigure}
\caption{
The asymptotic variance vs $k$ for two cases where the variance is minimized when $k \in (0, 1)$. That is, collecting a mix of confounders and mediators is better than collecting only confounders or mediators.
}
\label{fig:partial-cramer-rao}
\end{center}
\end{figure}

\noindent \textbf{Cramer-Rao Lower Bound \quad}
To compute the variance of the estimate of
$e = ac$, 
we compute the Cramer-Rao variance lower bound. 
We first compute the Fisher information matrix $\mathbf{I}$ 
as $\mathbf{I} = - \E \left[ \nabla^2_\theta \mathcal{LL} \right]$, where $\theta$ represents the eight model parameters.
Let $\widehat{e}$ be the MLE. Since regularity holds for our model (due to linearity and Gaussianity), the MLE is asymptotically normal. Using the Cramer-Rao theorem, for constant $k$, as $N \to \infty$, we have $\sqrt{N}(\widehat{e} - e) \overset{d}{\to} \mathcal{N}(0, V_e)$, where $V_e$ is a function of $\mathbf{I}^{-1}$.
The closed form expression for $V_e$ is given in Appendix \ref{sec:appendix-cramer-rao-bound}.

For any fixed $k \in (0, 1)$,
$(V_e - \AVar(\sqrt{P}\widehat{ac})_{\text{backdoor}}) < 0$ and $(V_e - \AVar(\sqrt{Q}\widehat{a}_f \widehat{c})) < 0$, where
$\AVar$ is asymptotic variance.
This shows that the combined estimator always has lower 
asymptotic variance than that of the backdoor and frontdoor estimators on the individual datasets. Moreover, we also find cases where the combined estimator outperforms both the backdoor and frontdoor estimators even when the total number of samples are the same.
That is, there exist model parameters such that $(V_e - \AVar(\sqrt{N}\widehat{ac})_{\text{backdoor}}) < 0$ and $(V_e - \AVar(\sqrt{N}\widehat{a}_f \widehat{c})) < 0$ for some $k \in (0, 1)$.
This means that is these cases, it is better to collect a mix of confounders and mediators rather than only collecting mediators or confounders.
Despite having access to the same number of samples, a mix of confounders and mediators can lead to lower variance.
This happens when the variances of the backdoor and frontdoor estimators are close to each other.
In Figure \ref{fig:partial-cramer-rao}, 
we present two examples of causal graphs 
where having a mix of confounders and mediators leads to the lowest asymptotic variance
(see Appendix \ref{sec:appendix-cramer-rao-comparison} 
for parameter values).

\noindent \textbf{The Maximum Likelihood Estimator \quad}
Computing an analytical solution for the model parameters 
that maximizes the log-likelihood turns out to be intractable. 
As a result, we update our estimated parameters 
to maximize the likelihood numerically.
The likelihood in Eq. \ref{eq:partial_log_likelihood} is non-convex. 
So we intialize the parameters using the two datasets (see Appendix \ref{sec:appendix-cramer-rao-param-init} for details) and run the Broyden–Fletcher–Goldfarb–Shanno (BFGS) algorithm \citep{fletcher2013practical} to maximize the likelihood.
In our experiments, 
the non-convexity of the likelihood 
never proved a practical problem.
When we find the global minimum, 
this estimator is optimal and dominates 
both the backdoor and frontdoor estimators.

%% file: sections/70-experiments.tex
\noindent \textbf{Synthetic Data  \quad}
To show that the empirical variance of the various estimators 
is close to the theoretical variance
(Table \ref{table:synthetic-theoretical-vs-empirical-variance}), we randomly initialize parameters and for each instance, we compute the Mean and Standard Deviation of Absolute Percentage Error
of theoretical variance as a predictor of empirical variance
(see Appendix \ref{sec:appendix-synthetic-experiments}). Next, we compare the estimators under different settings of the model parameters. Unless stated otherwise, the model parameter values we use for experiments are $a = 10, b = 4, c = 5, d = 5, \sigma^2_{u_w} = 1, \sigma^2_{u_x} = 1, \sigma^2_{u_m} = 1, \sigma^2_{u_y} = 1$.

\begin{figure}[t]
\begin{center}
\begin{subfigure}{0.48\columnwidth}
\includegraphics[width=0.99\columnwidth]{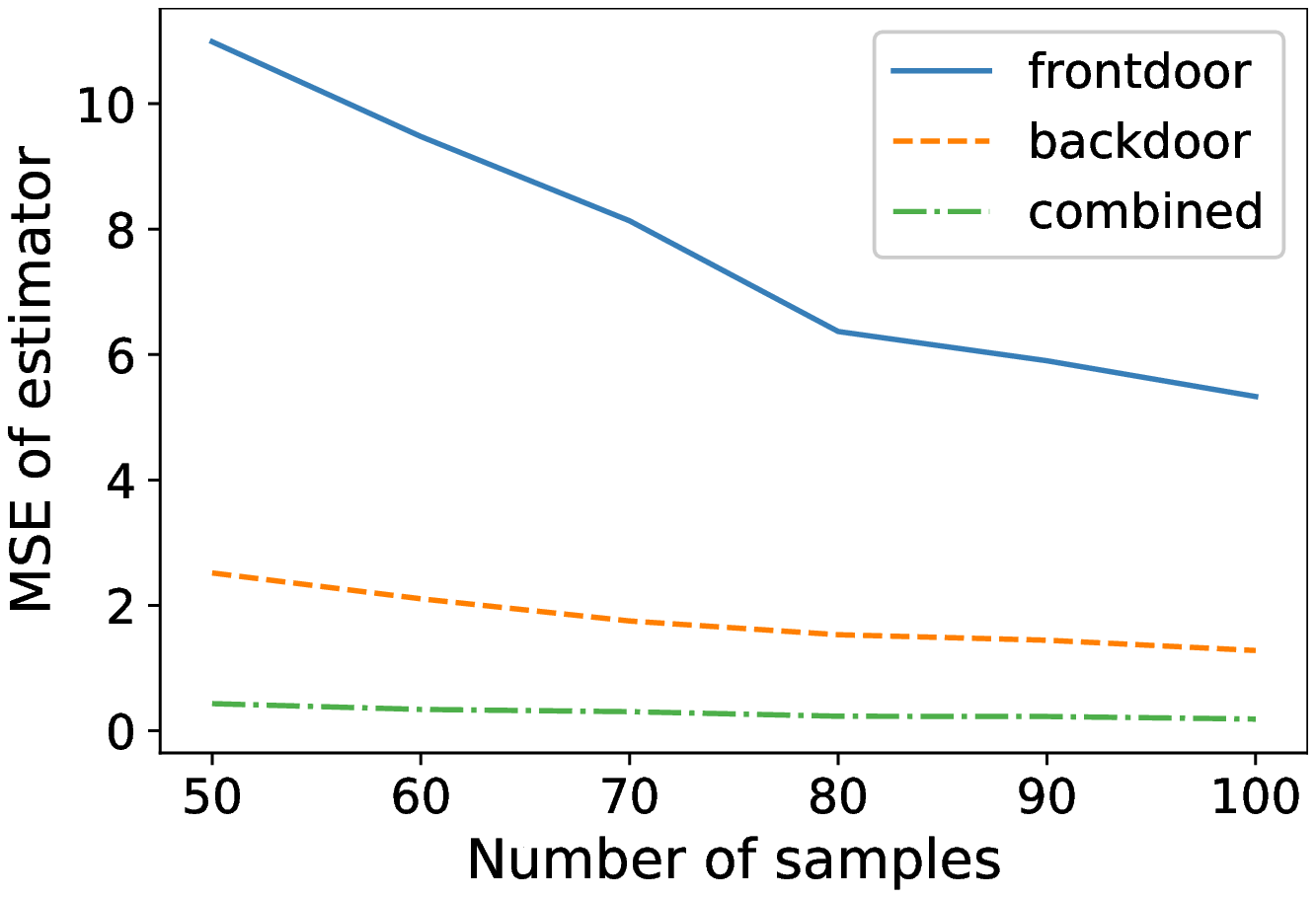}
\caption{Backdoor better}
\label{fig:frontdoor-backdoor-compare-backdoor-better}
\end{subfigure}
\begin{subfigure}{0.48\columnwidth}
\includegraphics[width=0.99\columnwidth]{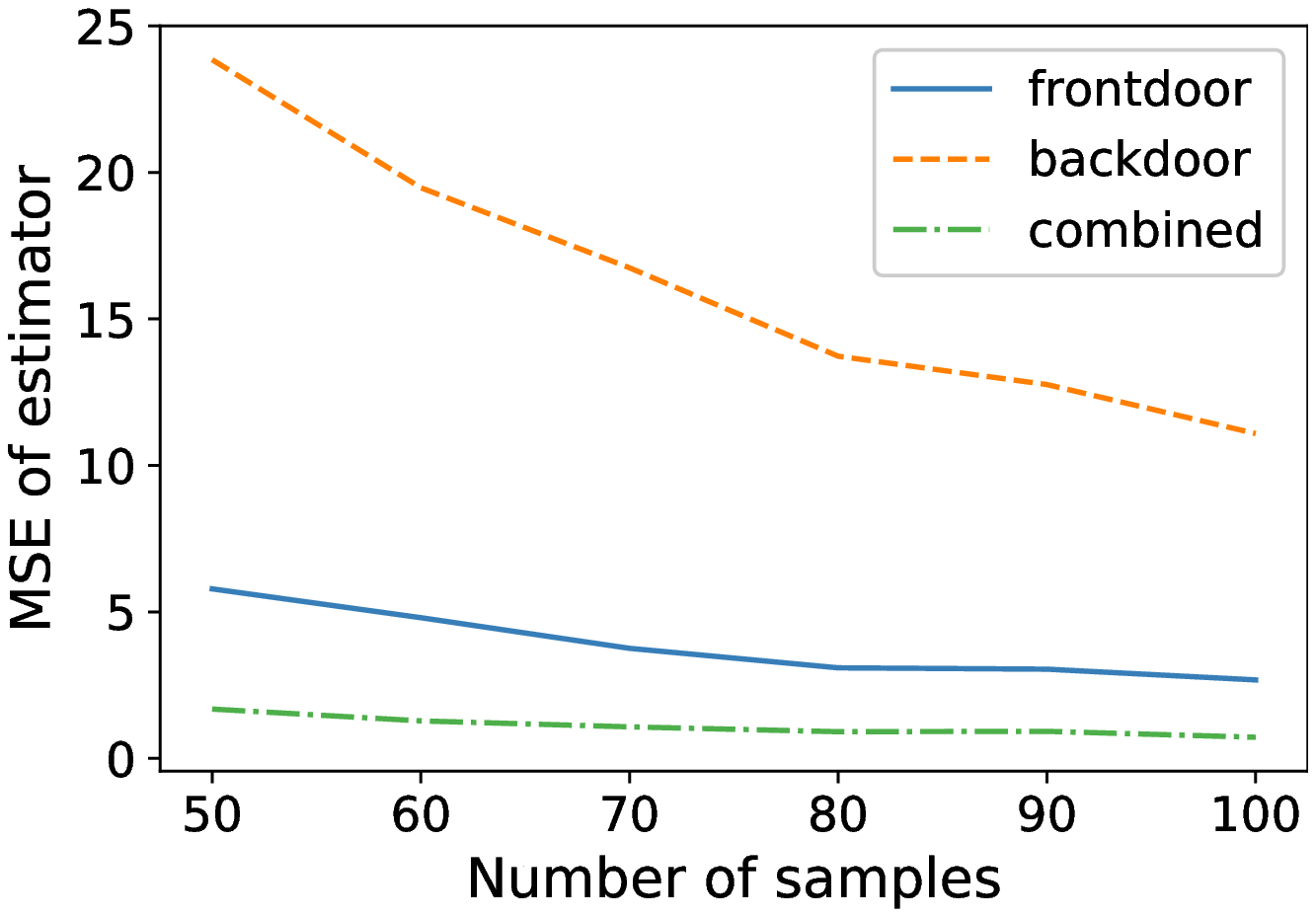}
\caption{Frontdoor better}
\label{fig:frontdoor-backdoor-compare-frontdoor-better}
\end{subfigure}
\end{center}
\caption{Comparison of MSE when confounders and mediators are observed simultaneously. Either of the backdoor or frontdoor estimators can dominate. The combined estimator is better than both.
}
\end{figure}

\begin{figure}[t]
\begin{center}
\end{center}
\begin{subfigure}{0.48\columnwidth}
\includegraphics[width=0.99\columnwidth]{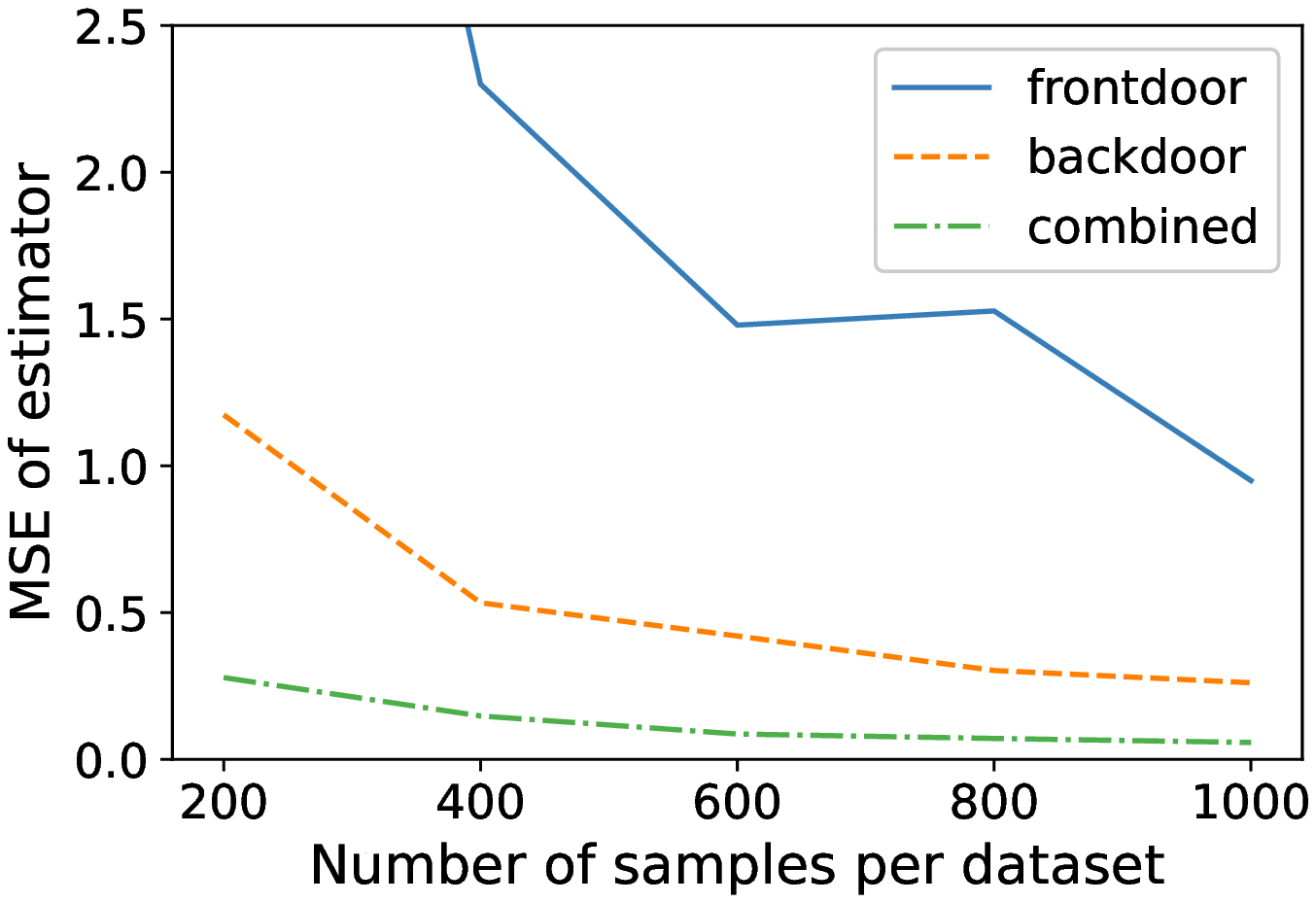}
\caption{}
\label{fig:partial-observe-compare-backdoor-better}
\end{subfigure}
\begin{subfigure}{0.48\columnwidth}
\includegraphics[width=0.99\columnwidth]{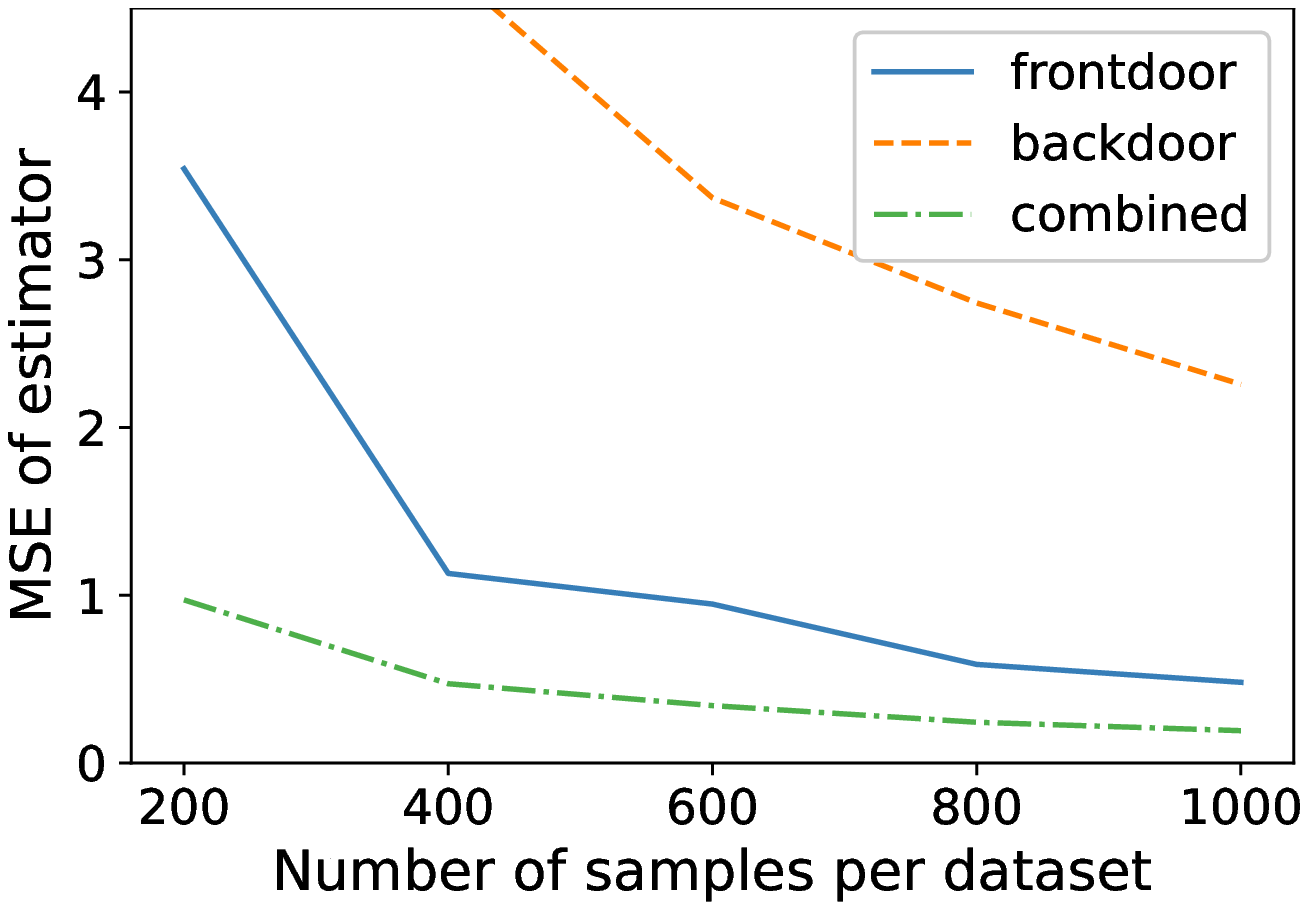}
\caption{}
\label{fig:partial-observe-compare-frontdoor-better}
\end{subfigure}
\caption{Comparison of MSE when confounders and mediators are observed in separate datasets. The combined estimator dominates in both cases.
}
\label{fig:partial-observe-compare}
\end{figure}

The quantity of interest is the causal effect $ac = 50$.
For Figure \ref{fig:frontdoor-backdoor-compare-backdoor-better}, we set $\{\sigma^2_{u_x} = 0.05, \sigma^2_{u_m} = 0.05\}$,
which makes the backdoor estimator better 
as predicted by Eq. \ref{eq:ratio-backdoor-to-frontdoor}. 
For Figure \ref{fig:frontdoor-backdoor-compare-frontdoor-better},
we set $\{\sigma^2_{u_w} = 2, \sigma^2_{u_x} = 0.01, \sigma^2_{u_m} = 0.1\}$ which makes the frontdoor estimator better as predicted by Eq. \ref{eq:ratio-backdoor-to-frontdoor}.
The plots in Figure \ref{fig:frontdoor-backdoor-compare-backdoor-better} and \ref{fig:frontdoor-backdoor-compare-frontdoor-better} corroborate these predictions at different sample sizes. Furthermore, the optimal combined estimator always outperforms 
both the backdoor and frontdoor estimators.

Next, we evaluate the procedure for combining datasets described in Section \ref{sec:partially-observed},
generating two datasets with equal numbers of samples. 
In the first, only $\{X, Y, W\}$ are observed. 
In the second, only $\{X, Y, M\}$ are observed. 
We set $\{\sigma^2_{u_x} = 0.05, \sigma^2_{u_m} = 0.05\}$,
which makes the backdoor estimator better
(Figure \ref{fig:partial-observe-compare-backdoor-better}),
and then set $\{\sigma^2_{u_w} = 2, \sigma^2_{u_x} = 0.01, \sigma^2_{u_m} = 0.1\}$,
which makes the frondoor estimator better
(Figure \ref{fig:partial-observe-compare-frontdoor-better}).
The plots show that the combined estimator 
has lower MSE than either for various sample sizes
(Figure \ref{fig:partial-observe-compare}),
supporting our theoretical claims.

\noindent \textbf{IHDP Dataset \quad}\label{sec:ihdp-results}
\citet{hill2011bayesian} constructed a dataset from 
the Infant Health and Development Program (IHDP). 
This semi-synthetic dataset has been used 
for benchmarking in causal inference \citep{shi2019adapting, shalit2017estimating}.
The dataset is
is based on a randomized experiment to measure
the effect of home visits from a specialist 
on future cognitive test scores of children.
The treatment is binary
and
the covariates contain both continuous and categorical variables
representing measurements on the child and their mother.
We use samples from the NPCI package \citep{Dorie2016}.
We converted the randomized data into an observational study
by removing a biased subset of the treated group.
This set contains $747$ samples with $25$ covariates.

We use the covariates and the treatment assignment 
from the real study. 
We use a procedure similar to \citet{hill2011bayesian} 
to simulate the mediator and the outcome. 
The mediator $M$ takes the form $M \sim \mathcal{N}(cX, \sigma^2_{u_m})$, where $X$ is the treatment. 
The response $Y$ takes the form $Y \sim \mathcal{N}(aM + w^\top \mathbf{b}, 1)$ where $w$ is the vector of standardized 
(zero mean and unit variance) covariates 
and values in the vector $\mathbf{b}$ are randomly sampled
(0, 1, 2, 3, 4) with probabilities (0.5, 0.2, 0.15, 0.1, 0.05). The ground truth causal effect is $c\times a$.

We evaluate our estimators and the four IF estimators: IF-Fulcher (IF-Fulc), IF-Restricted (IF-Restr), IF-Frontdoor (IF-FD), and IF-Rotnitzky (IF-Rotz) (Section \ref{sec:influence-function-estimators}).
We test the estimators on two settings of the model parameters (Table \ref{table:ihdp-jtpa-estimator-comparison}, the \emph{Complete} dataset setting). 
The MSE values are computed across 1000 instantiations of the dataset created 
by simulating the mediators and outcomes. 
We first evaluate the estimators on the complete dataset of 747 samples. We see that for Setting $1$ (S1): $a=10, c=5, \sigma_{u_m} = 1$, the backdoor estimator dominates the frontdoor estimator whereas
for Setting $2$ (S2): $a=10,c=1,\sigma_{u_m} = 2$,
the frontdoor estimator is better.
In both cases, the combined estimator (Section \ref{sec:combined-estimator}) outperforms both estimators. Furthermore, we see that IF-Restricted outperforms IF-Frontdoor, showing the value of leveraging the covariates. Moreover, IF-Restricted also outperforms IF-Fulcher, suggesting
that incorporating the additional model restrictions improves performance. Next, we randomly split the data into two sets, 
one with the confounder observed 
and the other with the mediator observed,
finding that the estimator that combines the datasets (Section \ref{sec:partially-observed})
outperforms the frontdoor and backdoor estimator
(Table \ref{table:ihdp-jtpa-estimator-comparison}, the \emph{Partial} dataset setting).
We compute the MSE over 1000 realizations of the dataset.

\begin{table}[t]
\caption{Mean Absolute Percentage Error of the
theoretical variance as a predictor of the empirical variance.
The values are reported as mean $\pm$ std. The \% error is small even for small sample sizes.}
\label{table:synthetic-theoretical-vs-empirical-variance}
\begin{center}
\begin{sc}
\begin{tabular}{lccccr}
\toprule
Estimator & $n=50$ & $n=100$ & $n=200$\\
\midrule
Backdoor & $0.36 \pm .3$ & $0.32 \pm .2$ & $0.34 \pm .1$\\
Frontdoor & $0.33 \pm .2$ & $0.30 \pm .2$ & $0.23 \pm .1$\\
Combined & $1.20 \pm 1.1$ & $0.97 \pm .6$ & $0.58 \pm .2$\\
\bottomrule
\end{tabular}
\end{sc}
\end{center}
\end{table}

\begin{table}
\caption{Results on the IHDP and JTPA datasets. The \textit{complete} (C) data setting is when $\{W, X, M, Y\}$ are observed and \textit{partial} (P) is with $\{X, M, Y\}$ and $\{W, X, Y\}$ observed in two separate datasets. }
\label{table:ihdp-jtpa-estimator-comparison}
\begin{center}
\begin{small}
\begin{sc}
\begin{tabular}{lcccll}
\toprule
 & & \multicolumn{2}{c}{IHDP MSE} & \multicolumn{2}{c}{JTPA} \\
Estimator & Data & S1 & S2 & Var & MSE\\
\midrule
\text{Backdoor} & \text{C} & 2.14 & 1.07 & NA & NA\\
\text{Frontdoor} & \text{C} & 1.97 & 2.81 & 40.9k & 75.3k\\
\text{Combined} & \text{C} & \textbf{1.78} & \textbf{0.93} & \textbf{33.1k} & \textbf{70.1k}\\
\text{IF-FD} & \text{C} & 4.24 & 2.07 & 46.6k & 77.9k\\
\text{IF-Restr} & \text{C} & \textbf{3.49} & \textbf{1.48} & \textbf{40.4k} & \textbf{42.1k}\\
\text{IF-Fulc} & \text{C} & 3.82 & 1.87 & 45.1k & 46.2k\\
\text{IF-Rotz} & \text{C} & 3.58 & 2.01 & NA & NA\\
\midrule
\text{Backdoor} & \text{P} & 5.44 & 2.43 & NA & NA\\
\text{Frontdoor} & \text{P} & 3.92 & 4.94 & \textbf{74.8k} & \textbf{115.1k} \\
\makecell[l]{Combined} & \text{P} & \textbf{2.97} & \textbf{1.62} & 79.5k & 123.1k\\
\bottomrule
\end{tabular}
\end{sc}
\end{small}
\end{center}
\end{table}

\noindent \textbf{National JTPA Study \quad}
The National Job Training Partnership Act (JTPA) Study evaluates the effect of a job training program on future earnings.
We use the dataset from \citet{glynnData2019}.
The binary treatment $X$ represents if a participant signed up for the program. The outcome $Y$ represents future earnings. The collected covariates (like race, study location, age) are the confounders $W$. 
The covariates contain both categorical and continuous variables.
There was non-compliance among the treated units. The binary mediator $M$ represents compliance, that is, whether the participant make use of JTPA services after signing up.
The study contained a randomized component which allowed us to compute the ground truth treatment effect, which was \textbf{$862.74$}. \citet{glynn2018front} showed that the backdoor estimator has high bias, suggesting that
there was unmeasured confounding, so we omit the backdoor estimator in our results.
They also justify the assumptions required for the frontdoor estimator and show that it works well for this study.

A comparison of the frontdoor estimator, the combined estimator (Section \ref{sec:combined-estimator}), IF-Restricted, IF-Frontdoor and IF-Fulcher (Section \ref{sec:influence-function-estimators}) is shown in Table \ref{table:ihdp-jtpa-estimator-comparison} (the ``C'' data setting). For IF-Restricted, we only use the $f(M|X,W)=f(M|X)$ restriction and do not use the $\E[Y|M,W,X]=\E[Y|M,W]$ restriction since it is not valid. We compute the variance and MSE using 1000 bootstrap iterations. The combined estimator has lower variance and MSE than the frontdoor estimator. 
IF-Restricted outperforms IF-Frontdoor, 
reinforcing the utility of combined estimators. 
Furthermore, IF-Restricted outperforms IF-Fulcher, showing that using model restrictions is valuable.
Next, we evaluate our procedure
for the partially-observed setting (Section \ref{sec:partially-observed}). 
We compute variance and MSE across $1000$ bootstrap iterations. At each iteration, we randomly split our dataset into two datasets of equal size, one with revealed confounders, one with revealed mediators.
The combined estimator does not outperform the frontdoor estimator (Table \ref{table:ihdp-jtpa-estimator-comparison}, the ``P'' data setting).
This is expected since the backdoor adjustment works poorly and 
the revealed-confounder data is unlikely to help.
Despite this, the combined estimator does not suffer too badly
and has low bias despite the required assumptions for one of the identification strategies not holding.

%% file: sections/80-conclusion.tex
In this paper, we studied over-identified graphs
with confounders and mediators,
showing that the two identification strategies 
can lead to estimators with arbitrarily different variances. 
We show that having access to both confounders and mediators
(either simultaneously or in separate datasets)
can give (unbounded) performance gains.
We also show that our results qualitatively apply to general non-linear settings.

\paragraph{Future Work}
We see several promising lines for future work, including
(i) extensions to more general graphs;
(ii) online data collection subject to some cost structure over the observations;
and (iii) leveraging overidentification to mitigate errors due to measurement and confounding.
Our experiments show the applicability of our methods in the frontdoor-backdoor graph,
with combined estimators yielding gains in both linear and non-linear settings. 
We expect these insights to extend to other over-identified settings (e.g. graphs with multiple instrumental variables, multiple confounders, etc.)
and we hope next to extend the results to more general over-identified causal graphs.
Additionally, we plan to analyze the online data collection setting.
Here, subject to budget constraints, a practitioner must choose 
which variables to observe at each time step.
This direction seems especially important in medical applications (where each test may be costly) 
and survey studies (with a cap on the number of questions).
Our current results suggest that the optimal strategy 
must depend on the model parameters. 
At each step, the revealed data will improve 
our estimates of the model parameters,
in turn impacting what we collect in the future.

One potential limitation of the method is that situations where there exist multiple valid identification formulas may be uncommon in practice, when finding a single source of identification can already be difficult. However, we believe that in reality, many identification approaches are often available,  
but members of the community find flaws in each of the proposed estimators. 
In these cases, with multiple imperfect estimators of the same causal effect,
we believe that overidentification might be leveraged 
to create robust combined estimators.

%% file: sections/90-appendix.tex
\onecolumn

\section{Brief review of OLS regression}

Since we use OLS regression for our results, 
we briefly review OLS estimators.
We consider the following setup:
\begin{align*}
    \mathbf{y} &= \mathbf{X} \mathbf{\beta} + \mathbf{e},
\end{align*}
where $\mathbf{y}$ and $\mathbf{e}$ are $n \times 1$ vectors, 
$\mathbf{X}$ is an $n \times d$ matrix of observations, and 
$\mathbf{\beta}$ is the $d \times 1$ coefficient vector that we want to estimate. 
If $\mathbf{e} \ind \mathbf{X}$ 
and $\mathbf{e} \sim \mathcal{N}(0, \sigma^2_e \mathbf{I_n})$, where $\mathbf{I_n}$ is the $n \times n$ identity matrix, then the OLS estimate of $\mathbf{\beta}$ is
\begin{align*}
    \widehat{\mathbf{\beta}} &= (\mathbf{X}^\top \mathbf{X})^{-1} \mathbf{X}^\top \mathbf{y} \nonumber \\
        &= \mathbf{\beta} + (\mathbf{X}^\top \mathbf{X})^{-1} \mathbf{X}^\top \mathbf{e},
\end{align*}
with $\mathbb{E}[\widehat{\mathbf{\beta}}] = \mathbf{\beta}$ and $\text{Var}(\widehat{\mathbf{\beta}}) = \sigma^2_e \mathbb{E}[(\mathbf{X}^\top \mathbf{X})^{-1}]$.
If each row $X_i$ of $\mathbf{X}$ is sampled from $X_i \overset{i.i.d.}{\sim} \mathcal{N}(0, \Sigma)$, then the distribution of $(\mathbf{X}^\top \mathbf{X})^{-1}$ is an Inverse-Wishart distribution. 
Then the variance of $\widehat{\mathbf{\beta}}$ is
\begin{align}\label{eq:appendix-inverse-wishart-var}
    \text{Var}(\widehat{\mathbf{\beta}}) = \frac{\sigma^2_e \Sigma^{-1}}{n - d - 1}.
\end{align}

\section{Covariance of \texorpdfstring{$\widehat{a}$}{a\_hat} and \texorpdfstring{$\widehat{c}$}{c\_hat}}

\subsection{Frontdoor estimator} \label{sec:appendix-frontdoor-covariance-zero}

We prove that $\text{Cov}(\widehat{a}_f, \widehat{c}) = 0$ for the frontdoor estimator. The expressions for $\widehat{a}_f$ and $\widehat{c}$ are
\begin{align}
    \widehat{c} &\begin{aligned}[t]
    &= \frac{\sum x_i m_i}{\sum x^2_i}\\
    &= c + \frac{\sum x_i u^m_i}{\sum x^2_i}
    \end{aligned} \label{eq:frontdoor_c_hat_expr}\\
    \widehat{a}_f &\begin{aligned}[t]
    &= \frac{\sum x^2_i \sum m_i y_i - \sum x_i m_i \sum x_i y_i}{\sum x^2_i \sum m^2_i - (\sum x_i m_i)^2}\\
    &= a + \frac{\sum x^2_i \sum m_i e_i - \sum x_i m_i \sum x_i e_i}{\sum x^2_i \sum m^2_i - (\sum x_i m_i)^2},
    \end{aligned} \label{eq:frontdoor_a_hat_expr}
\end{align}
where $e_i = -\frac{b}{d}u^x_i + u^y_i$. Using the fact the $(u^x, x)$ is bivariate normally distributed, we get
\begin{align}
    \E[e|x] &= \frac{b \sigma^2_{u_x}}{d (d^2 \sigma^2_{u_w} + \sigma^2_{u_x})} x \nonumber \\
        &= F x, \label{eq:appendix-e-expectation}
\end{align}
where $F =  \frac{b \sigma^2_{u_x}}{d (d^2 \sigma^2_{u_w} + \sigma^2_{u_x})}$. The covariance then is
\begin{align}
    \text{Cov}(\widehat{a}_f, \widehat{c}) &= \mathbb{E}[(\widehat{a}_f - a)(\widehat{c} - c)] \nonumber \\
    &= \mathbb{E}[\mathbb{E}[(\widehat{a}_f - a)(\widehat{c} - c)|x, m]] \nonumber \\
    &=\mathbb{E}\bigg[ \mathbb{E} \bigg[\left( \frac{\sum x^2_i \sum m_i e_i - \sum x_i m_i \sum m_i e_i}{\sum x^2_i \sum m^2_i - (\sum x_i m_i)^2} \right) \left( \frac{\sum x_i u^m_i}{\sum x^2_i} \right) \bigg|x,m \bigg] \bigg] \nonumber \\
    &= \mathbb{E}\bigg[ \left(\frac{\sum x^2_i \sum m_i \mathbb{E}[e_i|x] - \sum m_i x_i \sum x_i \mathbb{E}[e_i|x]}{\sum x^2_i \sum m^2_i - (\sum x_i m_i)^2}\right) \left(\frac{\sum u^m_i x_i}{\sum x^2_i}\right) \bigg] \label{eq:frontdoor-a-c-expanded} \\
    &= \mathbb{E}\bigg[ F \left(\frac{\sum x^2_i \sum m_i x_i - \sum m_i x_i \sum x^2_i}{\sum x^2_i \sum m^2_i - (\sum x_i m_i)^2}\right) \left(\frac{\sum u^m_i x_i}{\sum x^2_i}\right) \bigg] \nonumber \\
    &= 0, \nonumber
\end{align}
where in Eq. \ref{eq:frontdoor-a-c-expanded} we used the expression from Eq. \ref{eq:appendix-e-expectation}. Also, in Eq. \ref{eq:frontdoor-a-c-expanded}, we took $u^m_i$ out of the conditional expectation because $u^m_i$ is given $x_i$ and $m_i$ (because $u^m_i = m_i - c x_i$).

\subsection{Combined estimator} \label{sec:appendix-bothdoor-covariance-zero}

We prove that $\text{Cov}(\widehat{a}_c, \widehat{c}) = 0$ for the combined estimator from Section \ref{sec:combined-estimator}. The expressions for $\widehat{a}_c$ and $\widehat{c}$ are
\begin{align}
    \widehat{c} &\begin{aligned}[t]
    &= \frac{\sum x_i m_i}{\sum x^2_i}\\
    &= c + \frac{\sum x_i u^m_i}{\sum x^2_i}
    \end{aligned} \label{eq:bothdoor_c_hat_expr}\\
    \widehat{a}_c &\begin{aligned}[t]
    &= \frac{\sum w^2_i \sum m_i y_i - \sum w_i m_i \sum w_i y_i}{\sum w^2_i \sum m^2_i - (\sum w_i m_i)^2}\\
    &= a + \frac{\sum w^2_i \sum m_i u^y_i - \sum w_i m_i \sum w_i u^y_i}{\sum w^2_i \sum m^2_i - (\sum w_i m_i)^2}.
    \end{aligned} \label{eq:bothdoor_a_hat_expr}
\end{align}
The covariance is
\begin{align}
    \text{Cov}(\widehat{a}_c, \widehat{c}) &= \mathbb{E}[(\widehat{a}_c-a)(\widehat{c}-c)] \nonumber\\
    &= \mathbb{E}[\mathbb{E}[(\widehat{a}_c-a)(\widehat{c}-c)|x, m, w]] \nonumber \\ 
    &= \mathbb{E}\bigg[ \bigg(\frac{\sum w^2_i \sum m_i \mathbb{E}[u^y_i] - \sum m_i w_i \sum w_i \mathbb{E}[u^y_i]}{\sum w^2_i \sum m^2_i - (\sum w_i m_i)^2}\bigg) \bigg( \frac{\sum u^m_i x_i}{\sum x^2_i} \bigg) \bigg] \label{eq:bothdoor-a-c-expanded}\\
    &= 0 \nonumber,
\end{align}
where in \ref{eq:bothdoor-a-c-expanded} we used the fact that $\E[u^y_i] = 0$.

\section{Unbiasedness of the estimators}
\label{sec:appendix-unbiasedness}

\subsection{Backdoor estimator}
\label{sec:appendix-unbiasedness-backdoor}

Recall that for the backdoor estimator, we take the coefficient of $X$ in an OLS regression of $Y$ on $\{ X, W \}$. The outcome $y_i$ can be written as
\begin{align*}
    y_i = ac x_i + b w_i + a u^m_i + u^y_i.
\end{align*}
The error term $a u^m_i + u^y_i$ is independent of $(x_i, w_i)$. In this case, the OLS estimator is unbiased. Therefore, $\E[\widehat{ac}_{\text{backdoor}}] = ac$.

\subsection{Frontdoor estimator}
\label{sec:appendix-unbiasedness-frontdoor}
For the frontdoor estimator, we first compute $\widehat{c}$ by taking the coefficient of $X$ in an OLS regression of $M$ on $X$. The mediator $m_i$ can be written as
\begin{align*}
    m_i = c x_i + u^m_i.
\end{align*}
The error term $u^m_i$ is independent of $x_i$. In this case, the OLS estimator is unbiased and hence, $\E[\widehat{c}] = c$.

We then compute $\widehat{a}_f$ by taking the coefficient of $M$ in an OLS regression of $Y$ on $\{ M, X \}$. The outcome $y_i$ can be written as
\begin{align*}
    y_i &= a m_i + \frac{b}{d} x_i - \frac{b}{d} u^x_i + u^y.
\end{align*}
In this case, the error term $- \frac{b}{d} u^x_i + u^y$ is correlated with $x_i$. The expression for $\widehat{a}$ is given in Eq. \ref{eq:frontdoor_a_hat_expr}. The expectation $\E[\widehat{a}_f]$ is
\begin{align}
    \E[\widehat{a}_f] &= a + \E\left[ \frac{\sum x^2_i \sum m_i e_i - \sum x_i m_i \sum x_i e_i}{\sum x^2_i \sum m^2_i - (\sum x_i m_i)^2} \right] \nonumber \\
        &= a + \E\left[ \E\left[ \frac{\sum x^2_i \sum m_i e_i - \sum x_i m_i \sum x_i e_i}{\sum x^2_i \sum m^2_i - (\sum x_i m_i)^2} \bigg| x,m \right] \right] \nonumber \\
        &= a + \E\left[ \frac{\sum x^2_i \sum m_i \E[e_i|x] - \sum x_i m_i \sum x_i \E[e_i|x]}{\sum x^2_i \sum m^2_i - (\sum x_i m_i)^2} \right] \label{eq:appendix-expectation-e-used-1} \\
        &= a + \E\left[ \frac{\sum x^2_i \sum m_i (F x_i) - \sum x_i m_i \sum x_i (F x_i)}{\sum x^2_i \sum m^2_i - (\sum x_i m_i)^2} \right] \nonumber \\
        &= a, \nonumber
\end{align}
where, in Eq. \ref{eq:appendix-expectation-e-used-1}, the expression for $E[e_i|x]$ is taken from Eq. \ref{eq:appendix-e-expectation}. Using the fact that $\Cov(\widehat{a}_f, \widehat{c}) = 0$ (see proof in Appendix \ref{sec:appendix-frontdoor-covariance-zero}), we can see that the frontdoor estimator is unbiased as
\begin{align*}
    \E[\widehat{a}_f\widehat{c}] &= \E[\widehat{a}_f]\E[\widehat{c}] + \Cov(\widehat{a}_f, \widehat{c}) \\
        &= ac.
\end{align*}

\subsection{Combined estimator}
\label{sec:appendix-unbiasedness-combined}

In the combined estimator, the expression for $\widehat{c}$ is the same as for the frontdoor estimator. Therefore, as shown in Appendix \ref{sec:appendix-unbiasedness-frontdoor}, $\E[\widehat{c}] = c$. We compute $\widehat{a}$ by taking the coefficient of $M$ in an OLS regression of $Y$ on $\{ M, W \}$. The outcome $y_i$ can be written as
\begin{align*}
    y_i = a m_i + b w_i + u^y_i.
\end{align*}
The error term $u^y_i$ is independent of $(m_i, w_i)$. In this case, the OLS estimator is unbiased. Therefore, $\E[\widehat{a}_c] = a$. Using the fact that $\Cov(\widehat{a}_c, \widehat{c}) = 0$ (see proof Appendix \ref{sec:appendix-bothdoor-covariance-zero}), we can see that the combined estimator is unbiased as
\begin{align*}
    \E[\widehat{a}_c\widehat{c}] &= \E[\widehat{a}_c]\E[\widehat{c}] + \Cov(\widehat{a}_c, \widehat{c}) \\
        &= ac.
\end{align*}

\section{Variance results for the frontdoor, backdoor, and combined estimators}

\subsection{Backdoor estimator}\label{sec:appendix-variance-results-backdoor}

The outcome $y_i$ can be written as 
\begin{align*}
y_i = ac x_i + b w_i + a u^m_i + u^y_i.
\end{align*}
We estimate the causal effect $ac$ 
by taking the coefficient on $X$ in an OLS regression of $Y$ on $\{X, W\}$.
Let $\Sigma = \Cov([X, W])$. Using Eq.~\ref{eq:appendix-inverse-wishart-var}, 
the finite sample variance of the backdoor estimator is
\begin{align*}
    \Var(\widehat{ac})_{\text{backdoor}} &= \frac{\Var(a u^m + u^y) \left(\Sigma^{-1}\right)_{1,1}}{n - 3} \\
        &= \frac{a^2 \sigma^2_{u_m} + \sigma^2_{u_y}}{(n-3)\sigma^2_{u_x}}.
\end{align*}

OLS estimators are asymptotically normally for arbitrary error distributions (and hence, Gaussianity is not needed). Therefore, the asymptotic variance of the backdoor estimator is
\begin{align*}
    \lim_{n \to \infty} \Var(\sqrt{n}(\widehat{ac} - ac))_{\text{backdoor}} &= \Var(a u^m + u^y) \left(\Sigma^{-1}\right)_{1,1} = \frac{a^2 \sigma^2_{u_m} + \sigma^2_{u_y}}{\sigma^2_{u_x}}.
\end{align*}

\subsection{Frontdoor estimator}
\label{sec:appendix-variance-results-frontdoor}

\subsubsection{Variance of \texorpdfstring{$\widehat{c}$}{c\_hat}}

The regression of $M$ on $X$ can be written as $m_i = c x_i + u^m_i$. Let $\Sigma_c = \Var(X)$. Using Eq. \ref{eq:appendix-inverse-wishart-var}, $\Var(\widehat{c})$ is
\begin{align*}
    \Var(\widehat{c}) &= \frac{\Var(u^m) (\Sigma^{-1}_c)}{n-2}
        = \frac{\sigma^2_{u_m}}{(n-2)(d^2 \sigma^2_{u_w} + \sigma^2_{u_x})}.
\end{align*}

\subsubsection{Variance of \texorpdfstring{$\widehat{a}_f$}{a\_hat frontdoor}}

The regression of $Y$ on $\{M, X\}$ can be written as $y_i = a m_i + \frac{b}{d} x_i + e_i$, where $e_i = -\frac{b}{d} u^x_i + u^y_i$. 
In this case, the error $e_i$ is not independent of the regressor $x_i$. 
Using the fact that $(u^x, x)$ has a bivariate normal distribution, $\text{Var}(e|x)$ is
\begin{align}
    \text{Var}(e|x) &= \frac{b^2 \sigma^2_{u_w} \sigma^2_{u_x} + \sigma^2_{u_y} (d^2 \sigma^2_{u_w} + \sigma^2_{u_x})}{(d^2 \sigma^2_{u_w} + \sigma^2_{u_x})} \label{eq:appendix-var-e-given-x} \\
         &:= V_e \nonumber.
\end{align}
Note that $V_e$ is a constant and does not depend on $x$.
From Eqs. \ref{eq:frontdoor_c_hat_expr} and \ref{eq:frontdoor_a_hat_expr}, we know that
\begin{align*}
    \widehat{c} &= c + \frac{\sum x_i u^m_i}{\sum x^2_i} \\
    \widehat{a}_f &= a + \frac{\sum x^2_i \sum m_i e_i - \sum x_i m_i \sum x_i e_i}{\sum x^2_i \sum m^2_i - (\sum x_i m_i)^2}. 
\end{align*}
where $e_i = -\frac{b}{d}u^x_i + u^y_i$.
Let
\begin{align*}
    A &= \frac{\sum x^2_i \sum m_i e_i - \sum x_i m_i \sum x_i e_i}{\sum x^2_i \sum m^2_i - (\sum x_i m_i)^2} \\
    C &= \frac{\sum x_i u^m_i}{\sum x^2_i}.
\end{align*}

First, we derive the expression for $\Var(\widehat{a}_f)$ as follows,
\begin{align}
    \Var(\widehat{a}_f) &= \Var(a + A) \nonumber \\
        &= \Var(A) \nonumber \\
        &= \Var(\E[A|x, m]) + \E[\Var(A|x,m)] \nonumber \\
        &= \Var\left(\E\left[\frac{\sum x^2_i \sum m_i e_i - \sum x_i m_i \sum x_i e_i}{\sum x^2_i \sum m^2_i - (\sum x_i m_i)^2} \bigg|x, m\right]\right) + \E[\Var(A|x,m)] \nonumber \\
        &= \Var\left(\E\left[\frac{\sum x^2_i \sum m_i \E[e_i|x] - \sum x_i m_i \sum x_i \E[e_i|x]}{\sum x^2_i \sum m^2_i - (\sum x_i m_i)^2}\right]\right) + \E[\Var(A|x,m)] \label{eq:appendix-expectation-e-given-x-used-3} \\
        &= \Var\left(\E\left[\frac{\sum x^2_i \sum m_i (F x_i) - \sum x_i m_i \sum x_i (F x_i)}{\sum x^2_i \sum m^2_i - (\sum x_i m_i)^2}\right]\right) + \E[\Var(A|x,m)] \nonumber \\
        &= \E[\Var(A|x,m)] \nonumber \\
        &= \E\left[\Var\left(\frac{\sum x^2_i \sum m_i e_i - \sum x_i m_i \sum x_i e_i}{\sum x^2_i \sum m^2_i - (\sum x_i m_i)^2} \bigg|x,m\right)\right] \nonumber \\
        &= \E\left[ \frac{1}{\left( \sum x^2_i \sum m^2_i - (\sum x_i m_i)^2 \right)^2} \Var\left( \sum x^2_i \sum m_i e_i - \sum x_i m_i \sum x_i e_i \bigg|x,m\right)\right] \nonumber \\
        &= \E\left[ \frac{1}{\left( \sum x^2_i \sum m^2_i - (\sum x_i m_i)^2 \right)^2} \Var(e_i|x_i) \sum x^2_i \left( \sum x^2_i \sum m^2_i - (\sum x_i m_i)^2 \right) \right] \nonumber \\
        &= V_e \E\left[\frac{\sum x^2_i}{\sum x^2_i \sum m^2_i - (\sum x_i m_i)^2}\right] \nonumber \\
        &= V_e \E\left[D\right], \label{eq:appendix-variance-a-hat-intermediate}
\end{align}
where, in Eq. \ref{eq:appendix-expectation-e-given-x-used-3}, we used the result from Eq. \ref{eq:appendix-e-expectation}, and $D = \frac{\sum x^2_i}{\sum x^2_i \sum m^2_i - (\sum x_i m_i)^2}$. Using the fact that $D$ has the distribution of a marginal from an inverse Wishart-distributed matrix, that is, if the matrix $M \sim \mathcal{IW}(\Cov([M, X])^{-1}, n)$, then $D = M_{1,1}$, in Eq. \ref{eq:appendix-variance-a-hat-intermediate}, we get
\begin{align*}
    \Var(\widehat{a}_f) &= V_e \E\left[D\right] \\
        &= V_e \frac{\Cov([M, X])^{-1}_{1,1}}{n - 2 - 1} \\
        &=  \frac{b^2 \sigma^2_{u_w} \sigma^2_{u_x} + \sigma^2_{u_y} (d^2 \sigma^2_{u_w} + \sigma^2_{u_x})}{(n-3) (d^2 \sigma^2_{u_w} + \sigma^2_{u_x}) \sigma^2_{u_m}},
\end{align*}
where the expression for $V_e$ is taken from Eq. \ref{eq:appendix-var-e-given-x}.

\subsubsection{Covariance of \texorpdfstring{$\widehat{a}_f^2$}{a\_hat square frontdoor} and \texorpdfstring{$\widehat{c}^2$}{c\_hat square}}
\label{sec:appendix-finite-sample-frontdoor-covariance}

We prove that $\Cov(\widehat{a}_f^2, \widehat{c}^2) = \Var(\widehat{a}_f)\Var(\widehat{c})$. This covariance can be written as
\begin{align}
    \Cov(\widehat{a}_f^2, \widehat{c}^2) &= \E[(\widehat{a}_f^2 - \E[\widehat{a}_f^2])(\widehat{c}^2 - \E[\widehat{c^2}])] \nonumber \\
        &= \E[(\widehat{a}_f^2 - \Var(\widehat{a}_f) - \E^2[\widehat{a}_f])(\widehat{c}^2 - \Var(\widehat{c}) - \E^2[\widehat{c}])] \nonumber \\
        &= \E[\widehat{a}_f^2\widehat{c}^2] - \Var(\widehat{a}_f)\Var(\widehat{c}) - a^2 \Var(\widehat{c}) - c^2 \Var(\widehat{a}_f) - a^2 c^2 \label{eq:appendix-frontdoor-covariance-full}.
\end{align}
We can write $\E[\widehat{a}_f^2\widehat{c}^2]$ as
\begin{align}
    \E[\widehat{a}_f^2\widehat{c}^2] &= \E[(a + A)^2 (c + C)^2] \nonumber\\
        &= \E[a^2 c^2 + c^2 A^2 + a^2 C^2 + A^2 C^2 + 2 aA C^2 + 2cC A^2] \nonumber\\
        &= a^2 c^2 + c^2 \Var(\widehat{a}) + a^2 \Var(\widehat{c}) + \E[A^2 C^2] + \E[2 aA C^2] + \E[2cC A^2] \label{eq:appendix-frontdoor-expectation-full}.
\end{align}
Substituting the result from Eq. \ref{eq:appendix-frontdoor-expectation-full} in Eq. \ref{eq:appendix-frontdoor-covariance-full}, we get
\begin{align}
    \Cov(\widehat{a}_f^2, \widehat{c}^2) &= \E[A^2 C^2] + \E[2 aA C^2] + \E[2cC A^2] \label{eq:appendix-frontdoor-covariance-simplified}.
\end{align}
Now we expand each term in Eq. \ref{eq:appendix-frontdoor-covariance-simplified} separately. $\E[2 aA C^2]$ is
\begin{align}
    \E[2 aA C^2] &= 2 a \E\left[ \left(\frac{\sum x_i u^m_i}{\sum x^2_i} \right)^2 \left( \frac{\sum x^2_i \sum m_i e_i - \sum x_i m_i \sum x_i e_i}{\sum x^2_i \sum m^2_i - (\sum x_i m_i)^2} \right) \right] \nonumber \\
        &= 2 a \E\left[ \E\left[ \left(\frac{\sum x_i u^m_i}{\sum x^2_i} \right)^2 \left( \frac{\sum x^2_i \sum m_i e_i - \sum x_i m_i \sum x_i e_i}{\sum x^2_i \sum m^2_i - (\sum x_i m_i)^2} \right) \Bigg| {x, m} \right] \right] \nonumber \\
        &= 2 a \E\left[ \left(\frac{\sum x_i u^m_i}{\sum x^2_i} \right)^2 \left( \frac{\sum x^2_i \sum m_i \E[e_i|x] - \sum x_i m_i \sum x_i \E[e_i|x]}{\sum x^2_i \sum m^2_i - (\sum x_i m_i)^2} \right) \right] \label{eq:appendix-expectation-e-given-x-used} \\
        &= 2 a \E\left[ \left(\frac{\sum x_i u^m_i}{\sum x^2_i} \right)^2 \left( \frac{\sum x^2_i \sum m_i (F x_i) - \sum x_i m_i \sum x_i (F x_i)}{\sum x^2_i \sum m^2_i - (\sum x_i m_i)^2} \right) \right] \nonumber \\
        &= 0, \label{eq:appendix-frontdoor-covariance-zero-term}
\end{align}
where, in Eq. \ref{eq:appendix-expectation-e-given-x-used}, the expression for $\E[e|x]$ is taken from Eq. \ref{eq:appendix-e-expectation}. 

Next, we simplify $\E[2cC A^2]$ as
\begin{align}
    \E[&2cC A^2] \nonumber\\
    &= 2c \E\left[ \left(\frac{\sum x_i u^m_i}{\sum x^2_i} \right) \left( \frac{\sum x^2_i \sum m_i e_i - \sum x_i m_i \sum x_i e_i}{\sum x^2_i \sum m^2_i - (\sum x_i m_i)^2} \right)^2 \right] \nonumber \\
        &= 2c \E\left[ \E\left[ \left(\frac{\sum x_i u^m_i}{\sum x^2_i} \right) \left( \frac{\sum x^2_i \sum m_i e_i - \sum x_i m_i \sum x_i e_i}{\sum x^2_i \sum m^2_i - (\sum x_i m_i)^2} \right)^2 \Bigg| x,m \right] \right] \nonumber \\
        &= 2c \E\left[ \E\left[ \left(\frac{\sum x_i u^m_i}{\sum x^2_i} \right) \frac{(\sum x^2_i)^2 (\sum m_i e_i)^2 + (\sum x_i m_i)^2 (\sum x_i e_i)^2 - 2 \sum x^2_i \sum m_i e_i \sum x_i m_i \sum x_i e_i}{\left(\sum x^2_i \sum m^2_i - (\sum x_i m_i)^2 \right)^2} \Bigg| x,m \right] \right] \nonumber \\
        &= 2c \E\left[ \left(\frac{\sum x_i u^m_i}{\sum x^2_i} \right) \left( \frac{\Var(e|x) (\sum x^2_i)}{\sum x^2_i \sum m^2_i - (\sum x_i m_i)^2} + \frac{F^2 \left(2 (\sum x^2_i)^2 (\sum x_i m_i)^2 - 2 (\sum x^2_i)^2 (\sum x_i m_i)^2\right)}{\left(\sum x^2_i \sum m^2_i - (\sum x_i m_i)^2\right)^2} \right) \right] \nonumber \\
        &= 2c \E\left[ \left(\frac{\sum x_i u^m_i}{\sum x^2_i} \right) \left( \frac{\Var(e|x) (\sum x^2_i)}{\sum x^2_i \sum m^2_i - (\sum x_i m_i)^2}\right) \right] \nonumber \\
        &= 2c  V_e \E\left[ \left(\frac{\sum x_i u^m_i}{\sum x^2_i} \right) \left( \frac{ \sum x^2_i}{\sum x^2_i \sum m^2_i - (\sum x_i m_i)^2} \right) \right] \nonumber \\
        &= 2c  V_e \E\left[ \left(\widehat{c} - c \right) \left( \frac{ \sum x^2_i}{\sum x^2_i \sum m^2_i - (\sum x_i m_i)^2} \right) \right] \nonumber \\
        &= 2c  V_e \left( \E\left[ \widehat{c} \left( \frac{ \sum x^2_i}{\sum x^2_i \sum m^2_i - (\sum x_i m_i)^2} \right) \right] - c \E\left[ \left( \frac{ \sum x^2_i}{\sum x^2_i \sum m^2_i - (\sum x_i m_i)^2} \right) \right] \right) \nonumber \\
        &= 2c  V_e \left( \E\left[ \widehat{c} D \right] - c \E\left[ D \right] \right), \label{eq:appendix-frontdoor-covariance-first-term}
\end{align}
where $D = \frac{ \sum x^2_i}{\sum x^2_i \sum m^2_i - (\sum x_i m_i)^2}$.
Using the fact that $\widehat{c}$ and $D$ are independent of each other (see proof at the end of this section), we get
\begin{align}
    \E[\widehat{c} D] &= \E[\widehat{c}]\E[D] \nonumber \\
        &= c \E[D]. \label{eq:appendix-frontdoor-f-d-expectation}
\end{align}
Substituting the result from Eq. \ref{eq:appendix-frontdoor-f-d-expectation} in Eq. \ref{eq:appendix-frontdoor-covariance-first-term}, we get
\begin{align}
     \E[2cC A^2] &= 2c V_e \left( c \E\left[D \right] - c \E\left[ D \right] \right) \nonumber \\
        &= 0. \label{eq:appendix-frontdoor-covariance-first-upper-bound}
\end{align}
We proceed similarly to Eq. \ref{eq:appendix-frontdoor-covariance-first-term} to write $\E[A^2 C^2]$ as
\begin{align*}
    \E[A^2 C^2] &= V_e \E[C^2 D].
\end{align*}
Then we further simplify $\E[A^2 C^2]$ as
\begin{align}
    \E[A^2 C^2] &= V_e \E[C^2 D] \nonumber \\
        &= V_e \E[(\widehat{c} - c)^2 D] \nonumber \\
        &= V_e \left( \E\left[\widehat{c}^2\right] \E[D] - c^2 \E[D]  \right) \nonumber \\
        &= V_e \left( \Var(\widehat{c}) \E[D] + \E^2\left[\widehat{c}\right] \E[D] - c^2 \E[D]  \right) \nonumber \\
        &= V_e \Var(\widehat{c}) \E[D] \label{eq:appendix-inverse-wishart-1} \\
        &= V_e \Var(\widehat{c}) \frac{\Cov([M, X])^{-1}_{1,1}}{n - 2- 1} \nonumber \\
        &= \frac{V_e}{(n-3)\sigma^2_{u_m}} \Var(\widehat{c}) \label{eq:appendix-var-e-given-x-used-1} \\
        &= \Var(\widehat{a}_f) \Var(\widehat{c}), \label{eq:appendix-frontdoor-covariance-second-upper-bound}
\end{align}
where, in Eq. \ref{eq:appendix-inverse-wishart-1}, we used the fact that if the matrix $M \sim \mathcal{IW}(\Cov([M, X])^{-1}, n)$, then $D = M_{1,1}$ (that is, $D$ has the distribution of a marginal from an inverse Wishart-distributed matrix), and in Eq. \ref{eq:appendix-var-e-given-x-used-1}, the expression for $V_e$ is taken from Eq. \ref{eq:appendix-var-e-given-x}.

Substituting the results from Eqs. \ref{eq:appendix-frontdoor-covariance-zero-term}, \ref{eq:appendix-frontdoor-covariance-first-upper-bound}, and \ref{eq:appendix-frontdoor-covariance-second-upper-bound} in Eq. \ref{eq:appendix-frontdoor-covariance-simplified}, we get
\begin{align}
    \Cov(\widehat{a}_f^2, \widehat{c}^2) &= \Var(\widehat{a}_f) \Var(\widehat{c}). \label{eq:appendix-frontdoor-covariance-upper-bound}
\end{align}

\paragraph{Proof that $\widehat{c}$ and $D$ are independent.} Let $\Sigma$ be the following sample covariance matrix:
\begin{align*}
    \Sigma = \frac{1}{n} \begin{bmatrix}
        \sum m^2_i & \sum m_i x_i \\
        \sum m_i x_i & \sum x^2_i
    \end{bmatrix}.
\end{align*}
The distribution of $\Sigma$ is a Wishart distribution. That is, $\Sigma \sim \mathcal{W}(\Cov([M, X]), n)$. Then $(\Sigma_{1,1} - \Sigma_{1,2} \Sigma_{2,2}^{-1} \Sigma_{2,1})$ and $(\Sigma_{2,1}, \Sigma_{2,2})$ are independent \citep[Proposition 8.7]{eaton}. We can see that
\begin{align*}
    \Sigma_{1,1} - \Sigma_{1,2} \Sigma_{2,2}^{-1} \Sigma_{2,1} &= \frac{\sum m^2_i \sum x^2_i - \left(\sum x_i m_i\right)^2}{\sum x^2_i} \\
        &= \frac{1}{D}.
\end{align*}
Therefore, we get
\begin{align*}
    \frac{1}{D} &\ind \left(\sum x^2_i, \sum x_i m_i\right) \\
    \therefore\,\, D &\ind \left(\sum x^2_i, \sum x_i m_i\right) \\
    \therefore\,\, D &\ind \frac{\sum x_i m_i}{\sum x^2_i} \\
    \therefore\,\, D &\ind \widehat{c}.
\end{align*}

\subsubsection{Finite Sample Variance of \texorpdfstring{$\widehat{a}_f \widehat{c}$}{a\_hat frontdoor * c\_hat}}\label{sec:appendix-frontdoor-finite-sample-variance-result}

The variance of the product of two random variables can be written as
\begin{align}
    \Var(\widehat{a}_f\widehat{c}) &= \Cov(\widehat{a}_f^2, \widehat{c}^2) + (\Var(\widehat{a}_f) + \E^2[\widehat{a}_f])(\Var(\widehat{c}) + \E^2[\widehat{c}]) - (\Cov(\widehat{a}_f, \widehat{c}) + \E[\widehat{a}_f]\E[\widehat{c}])^2 \label{eq:appendix-frontdoor-variance-product-formula} \\
        &= \Cov(\widehat{a}_f^2, \widehat{c}^2) + (\Var(\widehat{a}_f) + a^2)(\Var(\widehat{c}) + c^2) - (\Cov(\widehat{a}_f, \widehat{c}) + ac)^2, \nonumber
\end{align}
where in Eq. \ref{eq:appendix-frontdoor-variance-product-formula} we used the facts that $\E[\widehat{a}_f] = a$, and $\E[\widehat{c}] = c$ (see Appendix \ref{sec:appendix-unbiasedness-frontdoor}). Using the facts that $\Cov(\widehat{a}_f^2, \widehat{c}^2) = \Var(\widehat{a}_f) \Var(\widehat{c})$ (from Eq. \ref{eq:appendix-frontdoor-covariance-second-upper-bound}) and $\Cov(\widehat{a}_f, \widehat{c}) = 0$ (from Appendix \ref{sec:appendix-frontdoor-covariance-zero}), we get
\begin{align*}
    \Var(\widehat{a}_f\widehat{c}) &= a^2 \Var(\widehat{c}) + c^2 \Var(\widehat{a}_f) + 2 \Var(\widehat{c}) \Var(\widehat{a}_f).
\end{align*}

\subsubsection{Asymptotic Variance of \texorpdfstring{$\widehat{a}_f \widehat{c}$}{a\_hat frontdoor * c\_hat}}\label{sec:appendix-frontdoor-asymptotic-variance-result}

Using asymptotic normality of OLS estimators, which does not require Gaussianity, we have
\begin{align*}
    \sqrt{n} \left( \begin{bmatrix}
    \widehat{a}_f \\
    \widehat{c}
    \end{bmatrix} -
    \begin{bmatrix}
    a \\
    c
    \end{bmatrix}\right) &\overset{d}{\to} \mathcal{N}\left(0, 
    \lim_{n \to \infty}  \begin{bmatrix}
    \Var_{\infty}(\widehat{a}_f) & \Cov(\sqrt{n} \widehat{a}_f, \widehat{c}) \\
    \Cov(\sqrt{n}\widehat{a}_f, \widehat{c}) & \Var_{\infty}(\widehat{c})
    \end{bmatrix}
    \right) \\
    \therefore\,\, \sqrt{n} \left( \begin{bmatrix}
    \widehat{a}_f \\
    \widehat{c}
    \end{bmatrix} -
    \begin{bmatrix}
    a \\
    c
    \end{bmatrix}\right) &\overset{d}{\to} \mathcal{N}\left(0, 
    \begin{bmatrix}
    \Var_{\infty}(\widehat{a}_f) & 0 \\
    0 & \Var_{\infty}(\widehat{c})
    \end{bmatrix}
    \right),
\end{align*}
where $\Var_{\infty}(\widehat{a}_f)$ and $\Var_{\infty}(\widehat{c})$ are the asymptotic variances of $\widehat{a}_f$ and $\widehat{c}$, respectively. The expressions for asymptotic variances are
\begin{align*}
    \Var_{\infty}(\widehat{a}_f) &= V_e \Cov([M, X])^{-1}_{1,1} =  \frac{b^2 \sigma^2_{u_w} \sigma^2_{u_x} + \sigma^2_{u_y} (d^2 \sigma^2_{u_w} + \sigma^2_{u_x})}{(d^2 \sigma^2_{u_w} + \sigma^2_{u_x}) \sigma^2_{u_m}}\\
    \Var_{\infty}(\widehat{c}) &= \Var(u^m) (\Sigma^{-1}_c)
        = \frac{\sigma^2_{u_m}}{d^2 \sigma^2_{u_w} + \sigma^2_{u_x}}.
\end{align*}
In order to compute the asymptotic variance of $\widehat{a}_f \widehat{c}$, we use the Delta method:
\begin{align*}
    \sqrt{n}(\widehat{a}_f \widehat{c} - ac) &\overset{d}{\to} \mathcal{N}\left(0,
    \begin{bmatrix}
    c & a
    \end{bmatrix}
    \begin{bmatrix}
    \Var_{\infty}(\widehat{a}_f) & 0 \\
    0 & \Var_{\infty}(\widehat{c})
    \end{bmatrix}
     \begin{bmatrix}
    c \\
    a
    \end{bmatrix}\right) \\
    \therefore\,\, \sqrt{n}(\widehat{a}_f \widehat{c} - ac) &\overset{d}{\to} \mathcal{N}\left(0,
     c^2 \Var_{\infty}(\widehat{a}_f) + a^2 \Var_{\infty}(\widehat{c}) \right).
\end{align*}

\subsection{Combined estimator}
\label{sec:appendix-combined-variance-results}

\subsubsection{Finite sample variance of \texorpdfstring{$\widehat{a}_c$}{a\_hat combined}} \label{sec:appendix-combined-a-hat-variance}

We can write the regression of $Y$ on $\{M, W\}$ 
as $y_i = a m_i + b w_i + u^y_i$. 
Let $\Sigma_{a_c} = \text{Cov}([M, W])$. 
Using Eq. \ref{eq:appendix-inverse-wishart-var}, 
we get
\begin{align*}
    \Var(\widehat{a}_c) &= \frac{\Var(u^y_i)(\Sigma^{-1}_{a_c})_{1,1}}{n-3} = \frac{\sigma^2_{u_y}}{(n-3)(c^2 \sigma^2_{u_x} + \sigma^2_{u_m})}.
\end{align*}

\subsubsection{Bounding the finite sample variance}\label{sec:appendix-combined-variance-finite-sample}

We first compute the lower bound of the combined estimator. Since the estimator is unbiased (see Appendix \ref{sec:appendix-bothdoor-covariance-zero}), we can apply the Cramer-Rao theorem to lower bound the finite sample variance.

Let the vector $\mathbf{s_i} = [x_i, y_i, w_i, m_i]$ denote the $i^{\text{th}}$ sample.
Since the data is multivariate Gaussian, 
the log-likelihood of the data is
\begin{align*}
    \mathcal{LL} = -\frac{n}{2} \left[\log{(\det{\Sigma})} + \Tr{(\widehat{\Sigma} \Sigma^{-1})}\right],
\end{align*}
where $\Sigma = \text{Cov}([X, Y, W, M])$
and $\widehat{\Sigma} = \frac{1}{n} \sum_{i=1}^{n} \mathbf{s_i} \mathbf{s_i}^\top$.
Let $e = ac$ and $\widehat{e} = \widehat{a}_c c$. Since we want to lower bound the variance of $\widehat{e}$, we reparameterize the log-likelihood by replacing $c$ with $e/a$ to simplify calculations. Next, we compute the Fisher Information Matrix for the eight model parameters:
\begin{align*}
    \mathbf{I} &= -E \begin{bmatrix}
    \frac{\partial^2 \mathcal{LL}}{\partial e^2} & \frac{\partial^2 \mathcal{LL}}{\partial e \partial a} & \frac{\partial^2 \mathcal{LL}}{\partial e \partial b} & \hdots & \frac{\partial^2 \mathcal{LL}}{\partial e \partial \sigma_{u_y}} \\
    \frac{\partial^2 \mathcal{LL}}{\partial a \partial e} & \frac{\partial^2 \mathcal{LL}}{\partial a^2} & \frac{\partial^2 \mathcal{LL}}{\partial a \partial b} & \hdots & \frac{\partial^2 \mathcal{LL}}{\partial a \partial \sigma_{u_y}} \\
    \vdots & \vdots & \vdots & \ddots & \vdots \\
    \frac{\partial^2 \mathcal{LL}}{\partial \sigma_{u_y} \partial e} & \frac{\partial^2 \mathcal{LL}}{\partial \sigma_{u_y} \partial a} & \frac{\partial^2 \mathcal{LL}}{\partial \sigma_{u_y} \partial b} & \hdots & \frac{\partial^2 \mathcal{LL}}{\partial^2 \sigma_{u_y}}
    \end{bmatrix}.
\end{align*}
Therefore, using the Cramer-Rao theorem, we have
\begin{align*}
    \Var(\widehat{e}) &= \Var(\widehat{a}_c c) \\
        &\geq (\mathbf{I}^{-1})_{1,1} \\
        &= \frac{1}{n} \left( \frac{c^2 \sigma^2_{u_y}}{c^2 \sigma^2_{u_x} + \sigma^2_{u_m}} + \frac{a^2 \sigma^2_{u_m}}{d^2 \sigma^2_{u_w} + \sigma^2_{u_x}} \right).
\end{align*}

Next, we compute a finite sample upper bound for $\Cov(\widehat{a}_c^2, \widehat{c}^2)$. We derive this in a similar manner as the frontdoor estimator in Appendix \ref{sec:appendix-finite-sample-frontdoor-covariance}.
From Eqs. \ref{eq:bothdoor_c_hat_expr} and \ref{eq:bothdoor_a_hat_expr}, we know that
\begin{align*}
    \widehat{a}_c &= a + \frac{\sum w^2_i \sum m_i u^y_i - \sum w_i m_i \sum w_i u^y_i}{\sum w^2_i \sum m^2_i - (\sum w_i m_i)^2}\\
    \widehat{c} &= c + \frac{\sum x_i u^m_i}{\sum x^2_i}.
\end{align*}
Let
\begin{align*}
    A &= \frac{\sum w^2_i \sum m_i u^y_i - \sum w_i m_i \sum w_i u^y_i}{\sum w^2_i \sum m^2_i - (\sum w_i m_i)^2} \\
    C &= \frac{\sum x_i u^m_i}{\sum x^2_i}.
\end{align*}
Then, similarly to Eq. \ref{eq:appendix-frontdoor-covariance-simplified}, we get
\begin{align}
    \Cov(\widehat{a}_c^2, \widehat{c}^2) &= \E[A^2 C^2] + \E[2 aA C^2] + \E[2cC A^2]. \label{eq:appendix-combined-covariance-simplified}
\end{align}
Now we simplify each term in Eq. \ref{eq:appendix-combined-covariance-simplified} separately. $\E[2 aA C^2]$ can be simplified as
\begin{align}
    \E[2 aA C^2] &= 2 a \E \left[ \left( \frac{\sum x_i u^m_i}{\sum x^2_i} \right)^2 \left( \frac{\sum w^2_i \sum m_i u^y_i - \sum w_i m_i \sum w_i u^y_i}{\sum w^2_i \sum m^2_i - (\sum w_i m_i)^2} \right) \right] \nonumber \\
        &= \E\left[ \E\left[ \left( \frac{\sum x_i u^m_i}{\sum x^2_i} \right)^2 \left( \frac{\sum w^2_i \sum m_i u^y_i - \sum w_i m_i \sum w_i u^y_i}{\sum w^2_i \sum m^2_i - (\sum w_i m_i)^2} \right) \Bigg|x,m,w \right] \right] \nonumber \\
        &= \E\left[ \left( \frac{\sum x_i u^m_i}{\sum x^2_i} \right)^2 \left( \frac{\sum w^2_i \sum m_i \E[u^y_i] - \sum w_i m_i \sum w_i \E[u^y_i]}{\sum w^2_i \sum m^2_i - (\sum w_i m_i)^2} \right) \right] \label{eq:appendix-combined-covariance-zero-intermediate} \\
        &= 0, \label{eq:appendix-combined-covariance-zero-term}
\end{align}
where, in Eq. \ref{eq:appendix-combined-covariance-zero-intermediate}, we used the fact that $\E[u^y] = 0$.

Next, we simplify $\E[2cC A^2]$ as
\begin{align}
    \E[2cC A^2] &= 2c \E\left[ \left( \frac{\sum x_i u^m_i}{\sum x^2_i} \right) \left( \frac{\sum w^2_i \sum m_i u^y_i - \sum w_i m_i \sum w_i u^y_i}{\sum w^2_i \sum m^2_i - (\sum w_i m_i)^2} \right)^2 \right] \nonumber \\
        &= 2c \E\left[ \E\left[ \left( \frac{\sum x_i u^m_i}{\sum x^2_i} \right) \left( \frac{\sum w^2_i \sum m_i u^y_i - \sum w_i m_i \sum w_i u^y_i}{\sum w^2_i \sum m^2_i - (\sum w_i m_i)^2} \right)^2 \Bigg| x,m,w \right] \right] \nonumber \\
        &= 2c \E\left[ \left( \frac{\sum x_i u^m_i}{\sum x^2_i} \right) \Var(u^y) \left( \frac{\sum w^2_i}{\sum w^2_i \sum m^2_i - (\sum w_i m_i)^2} \right) \right] \nonumber \\
        &= 2c \sigma^2_{u_y} \E\left[ C D \right], \label{eq:appendix-combined-covariance-first-term}
\end{align}
where $D = \frac{\sum w^2_i}{\sum w^2_i \sum m^2_i - (\sum w_i m_i)^2}$. We can upper bound the expression in Eq. \ref{eq:appendix-combined-covariance-first-term} as
\begin{align}
    \E[2cC A^2] &= 2c \sigma^2_{u_y} \E\left[ C D \right] \nonumber \\
        &\leq 2 |c| \sigma^2_{u_y} \E\left[ C D \right] \label{eq:appendix-cauchy-ineq-1} \\
        &\leq 2 |c| \sigma^2_{u_y} \sqrt{\E[C^2] \E[D^2]} \nonumber \\
        &= 2 |c| \sigma^2_{u_y} \sqrt{\Var(\widehat{c}) (\Var(D) + \E[D]^2)} \label{eq:appendix-d-inv-wishart-1} \\
        &= 2 |c| \sigma^2_{u_y} \sqrt{\Var(\widehat{c}) \left(\frac{2 \left[(\Cov([M, W]))^{-1}_{1,1}\right]^2}{(n-2-1)^2 (n-2-3)} + \left( \frac{(\Cov([M, W]))^{-1}_{1,1}}{n-2-1} \right)^2 \right)} \nonumber \\
        &= 2 |c| \sigma^2_{u_y} \sqrt{\Var(\widehat{c})} \sqrt{ \frac{2}{(n-3)^2 (n-5) (c^2 \sigma^2_{u_x} + \sigma^2_{u_m})^2 } + \frac{1}{(n-3)^2 (c^2 \sigma^2_{u_x} + \sigma^2_{u_m})^2}} \nonumber \\
        &= 2 |c| \frac{\sigma^2_{u_y}}{(n-3) (c^2 \sigma^2_{u_x} + \sigma^2_{u_m})} \sqrt{\Var(\widehat{c})} \sqrt{\frac{n-3}{n-5}} \nonumber \\
        &= 2 |c| \Var(\widehat{a}) \sqrt{\Var(\widehat{c})} \sqrt{\frac{n-3}{n-5}}, \label{eq:appendix-combined-covariance-first-upper-bound}
\end{align}
where, in Eq. \ref{eq:appendix-cauchy-ineq-1}, we used the Cauchy–Schwarz inequality, and in Eq. \ref{eq:appendix-d-inv-wishart-1}, we used the fact that if the matrix $M \sim \mathcal{IW}(\Cov([M, W])^{-1}, n)$, then $D = M_{1,1}$ (that is, $D$ has the distribution of a marginal from an inverse Wishart-distributed matrix).

Similarly to Eq. \ref{eq:appendix-combined-covariance-first-term}, we simplify $\E[A^2 C^2]$ as
\begin{align}
    \E[A^2 C^2] &= \sigma^2_{u_y} \E[C^2 D]. \label{eq:appendix-combined-covariance-second-term}
\end{align}
The expression in Eq. \ref{eq:appendix-combined-covariance-second-term} can be upper bounded using the Cauchy-Schwarz inequality as
\begin{align}
    \E[A^2 C^2] &= \sigma^2_{u_y} \E[C^2 D] \nonumber \\
        &\leq \sigma^2_{u_y} \sqrt{\E[C^4] \E[D^2]} \nonumber \\
        &= \sigma^2_{u_y} \sqrt{\E[C^4] (\Var(D) + \E^2[D])} \nonumber \\
        &= \sigma^2_{u_y} \sqrt{\E[C^4] \left(\frac{2 \left[(\Cov([M, W]))^{-1}_{1,1}\right]^2}{(n-2-1)^2 (n-2-3)} + \left( \frac{(\Cov([M, W]))^{-1}_{1,1}}{n-2-1} \right)^2 \right)} \nonumber \\
        &= \frac{\sigma^2_{u_y}}{(n-3) (c^2 \sigma^2_{u_x} + \sigma^2_{u_m})} \sqrt{\frac{n-3}{n-5}} \sqrt{\E[C^4]} \nonumber \\
        &= \Var(\widehat{a}_c) \sqrt{\frac{n-3}{n-5}} \sqrt{\E[C^4]}. \label{eq:appendix-combined-second-term-interm}
\end{align}
We can simplify $E[C^4]$ as follows,
\begin{align}
    E[C^4] &= \E\left[ \left( \frac{\sum x_i u^m_i}{\sum x^2_i} \right)^4 \right] \nonumber \\
        &= \E\left[ \E\left[ \left( \frac{\sum x_i u^m_i}{\sum x^2_i} \right)^4 \bigg| x \right] \right] \nonumber \\
        &= \E\left[ \frac{1}{\left( \sum x^2_i \right)^4} \E\left[ \left( \sum x_i u^m_i \right)^4 \bigg| x \right] \right] \nonumber \\
        &= \E\left[ \frac{1}{\left( \sum x^2_i \right)^4} \left\{ \Var\left( \left( \sum x_i u^m_i \right)^2 \big| x \right) + \E\left[ \left( \sum x_i u^m_i \right)^2 \big| x \right]^2 \right\} \right] \nonumber \\
        &= \E\left[ \frac{1}{\left( \sum x^2_i \right)^4} \left\{ \Var\left( \left( \sum x_i u^m_i \right)^2 \big| x \right) + \sigma^4_{u_m} \left( \sum x^2_i \right)^2 \right\} \right] \nonumber \\
        &= \E\left[ \frac{1}{\left( \sum x^2_i \right)^4} \left\{ \Var\left( \sigma^2_{u_m} \sum x^2_i \frac{\left( \sum x_i u^m_i \right)^2}{\sigma^2_{u_m} \sum x^2_i} \bigg| x \right) + \sigma^4_{u_m} \left( \sum x^2_i \right)^2 \right\} \right] \nonumber \\
        &= \E\left[ \frac{1}{\left( \sum x^2_i \right)^4} \left\{ \sigma^4_{u_m} \left(\sum x^2_i \right)^2 \Var\left( \frac{\left( \sum x_i u^m_i \right)^2}{\sigma^2_{u_m} \sum x^2_i} \bigg| x \right) + \sigma^4_{u_m} \left( \sum x^2_i \right)^2 \right\} \right] \label{eq:appendix-chi-square-1} \\
        &= \E\left[ \frac{1}{\left( \sum x^2_i \right)^4} \left\{ \sigma^4_{u_m} \left(\sum x^2_i \right)^2 2 + \sigma^4_{u_m} \left( \sum x^2_i \right)^2 \right\} \right] \nonumber \\
        &= \E\left[ \frac{1}{\left( \sum x^2_i \right)^4} \left\{3 \sigma^4_{u_m} \left( \sum x^2_i \right)^2 \right\} \right] \nonumber \\
        &= 3 \sigma^4_{u_m} \E\left[ \frac{1}{\left( \sum x^2_i \right)^2} \right] \nonumber \\
        &= 3 \sigma^4_{u_m} \left[ \Var\left( \frac{1}{\sum x^2_i} \right) + \E\left[ \frac{1}{\sum x^2_i} \right]^2 \right] \label{eq:appendix-scaled-inv-chi} \\
        &= 3 \sigma^4_{u_m} \left[ \frac{2}{(n-2)^2 (n-4) (d^2 \sigma^2_{u_w} + \sigma^2_{u_x})^2} + \frac{1}{(n-2)^2 (d^2 \sigma^2_{u_w} + \sigma^2_{u_x})^2} \right] \nonumber \\
        &= 3 \frac{\sigma^4_{u_m}}{(n-2)^2 (d^2 \sigma^2_{u_w} + \sigma^2_{u_x})^2} \left[ \frac{n-2}{n-4} \right] \nonumber \\
        &= 3 \left(\Var(\widehat{c}^2)\right)^2 \left[ \frac{n-2}{n-4} \right], \label{eq:appendix-eq-c-power-4}
\end{align}
where, in Eq. \ref{eq:appendix-chi-square-1}, we used the fact that $\frac{\left( \sum x_i u^m_i \right)^2}{\sigma^2_{u_m} \sum x^2_i} \bigg| x$ has a Chi-squared distribution, that is, $\frac{\left( \sum x_i u^m_i \right)^2}{\sigma^2_{u_m} \sum x^2_i} \bigg| x \sim \chi^2(1)$, and in Eq. \ref{eq:appendix-scaled-inv-chi}, we used the fact that $\frac{1}{\sum x^2_i}$ has a scaled inverse Chi-squared distribution, that is, $\frac{1}{\sum x^2_i} \sim \text{Scale-inv-}\chi^2\left(n, \frac{(d^2 \sigma^2_{u_w} + \sigma^2_{u_x})^2}{n}\right)$.

Substituting the result from Eq. \ref{eq:appendix-eq-c-power-4} in Eq. \ref{eq:appendix-combined-second-term-interm}, we get
\begin{align}
    \E[A^2 C^2] &\leq \Var(\widehat{a}_c) \Var(\widehat{c}) \sqrt{\frac{3 (n-3)(n-2)}{(n-5)(n-4)}}. \label{eq:appendix-combined-covariance-second-upper-bound}
\end{align}

Substituting the results from Eqs. \ref{eq:appendix-combined-covariance-zero-term}, \ref{eq:appendix-combined-covariance-first-upper-bound}, and \ref{eq:appendix-combined-covariance-second-upper-bound} in Eq. \ref{eq:appendix-combined-covariance-simplified}, we get
\begin{align*}
    \Cov(\widehat{a}_c^2, \widehat{c}^2) \leq \sqrt{\frac{n-3}{n-5}} \left( 2 |c| \Var(\widehat{a}_c) \sqrt{\Var(\widehat{c})} + \sqrt{3} \sqrt{\frac{n-2}{n-4}} \Var(\widehat{a}_c) \Var(\widehat{c}) \right).
\end{align*}
The variance of the product of two random variables can be written as
\begin{align*}
    \Var(\widehat{a}_c\widehat{c}) &= \Cov(\widehat{a}_c^2, \widehat{c}^2) + (\Var(\widehat{a}_c) + \E^2[\widehat{a}_c])(\Var(\widehat{c}) + \E^2[\widehat{c}]) - (\Cov(\widehat{a}_c, \widehat{c}) + \E[\widehat{a}_c]\E[\widehat{c}])^2 \\
        &= \Cov(\widehat{a}_c^2, \widehat{c}^2) + (\Var(\widehat{a}_c) + a^2)(\Var(\widehat{c}) + c^2) - (\Cov(\widehat{a}_c, \widehat{c}) + ac)^2,
\end{align*}
where we used the facts that $\E[\widehat{a}_c] = a$, and $\E[\widehat{c}] = c$ (see Appendix \ref{sec:appendix-unbiasedness-combined}). Using the fact that $\Cov(\widehat{a}_c, c) = 0$ (see Appendix \ref{sec:appendix-bothdoor-covariance-zero}) and the upper bound for $\Cov(\widehat{a}_c^2, \widehat{c}^2)$, we get
\begin{align*}
    \Var(&\widehat{a}_c \widehat{c}) \leq \\
        & c^2 \Var(\widehat{a}_c) + a^2 \Var(\widehat{c}) + \sqrt{\frac{n-3}{n-5}} \left( 2 |c| \Var(\widehat{a}_c) \sqrt{\Var(\widehat{c})} + \sqrt{3}\sqrt{\frac{n-2}{n-4}} \Var(\widehat{a}_c) \Var(\widehat{c}) \right).
\end{align*}

\subsubsection{Asymptotic variance}\label{sec:appendix-combined-asymptotic-variance}

Using asymptotic normality of OLS estimators, which does not require Gaussianity, we have
\begin{align*}
    \sqrt{n} \left( \begin{bmatrix}
    \widehat{a}_c \\
    \widehat{c}
    \end{bmatrix} -
    \begin{bmatrix}
    a \\
    c
    \end{bmatrix}\right) &\overset{d}{\to} \mathcal{N}\left(0, 
    \lim_{n \to \infty} \begin{bmatrix}
    \Var_{\infty}(\widehat{a}_c) & \Cov(\sqrt{n} \widehat{a}_c, \widehat{c}) \\
    \Cov(\sqrt{n} \widehat{a}_c, \widehat{c}) & \Var_{\infty}(\widehat{c})
    \end{bmatrix}
    \right) \\
    \therefore\,\, \sqrt{n} \left( \begin{bmatrix}
    \widehat{a}_c \\
    \widehat{c}
    \end{bmatrix} -
    \begin{bmatrix}
    a \\
    c
    \end{bmatrix}\right) &\overset{d}{\to} \mathcal{N}\left(0, 
    \begin{bmatrix}
    \Var_{\infty}(\widehat{a}_c) & 0 \\
    0 & \Var_{\infty}(\widehat{c})
    \end{bmatrix}
    \right),
\end{align*}
where $\Var_{\infty}(\widehat{a}_c)$ and $\Var_{\infty}(\widehat{c})$ are the asymptotic variances of $\widehat{a}_c$ and $\widehat{c}$, respectively. The expressions for the asymptotic variances are
\begin{align*}
    \Var_{\infty}(\widehat{a}_c) &= \Var(u^y_i)(\Sigma^{-1}_{a_c})_{1,1} = \frac{\sigma^2_{u_y}}{c^2 \sigma^2_{u_x} + \sigma^2_{u_m}} \\
    \Var_{\infty}(\widehat{c}) &= \Var(u^m) (\Sigma^{-1}_c)
        = \frac{\sigma^2_{u_m}}{d^2 \sigma^2_{u_w} + \sigma^2_{u_x}}.
\end{align*}
In order to compute the asymptotic variance of $\widehat{a}_c \widehat{c}$, we use the Delta method:
\begin{align*}
    \sqrt{n}(\widehat{a}_c \widehat{c} - ac) &\overset{d}{\to} \mathcal{N}\left(0,
     \begin{bmatrix}
    c & a
    \end{bmatrix}
    \begin{bmatrix}
    \Var_{\infty}(\widehat{a}_c) & 0 \\
    0 & \Var_{\infty}(\widehat{c})
    \end{bmatrix}
     \begin{bmatrix}
    c \\
    a
    \end{bmatrix}\right) \\
    \implies \sqrt{n}(\widehat{a}_c \widehat{c} - ac) &\overset{d}{\to} \mathcal{N}\left(0,
     c^2 \Var_{\infty}(\widehat{a}_c) + a^2 \Var_{\infty}(\widehat{c}) \right).
\end{align*}

\section{Comparison of combined estimator with backdoor and frontdoor estimators}\label{sec:appendix-combined-better}

In this section, we provide more details on the comparison of the combined estimator presented in \ref{sec:combined-estimator} to the backdoor and frontdoor estimators.

\subsection{Comparison with the backdoor estimator} \label{sec:appendix-combined-better-than-backdoor}

In Section \ref{sec:combined-estimator-dominates-better-of-both}, we made the claim that
\begin{align*}
    \exists N, \text{s.t.}, \forall n > N, \Var(\widehat{a}_c \widehat{c}) \leq \Var(\widehat{ac})_{\text{backdoor}}.
\end{align*}
In this case, by comparing Eqs. \ref{eq:backdoor_variance} and \ref{eq:combined-finite-sample-upper-bound}, we have
\begin{align*}
    N = \frac{2 \left( \sigma^4_{u_x} F + d^2 \sigma^2_{u_w} (\sigma^2_{u_m}D + \sigma^2_{u_x}F) + c^2 \sigma^6_{u_x} \sqrt{ c^2 \sigma^6_{u_x} D^2 (F + 2\sqrt{3}\sigma^2_{u_m})} \right)}{\sigma^2_{u_m} D^2},
\end{align*}
where $D = d^2 \sigma^2_{u_w} + \sigma^2_{u_x}$, $E = c^2 \sigma^2_{u_x} + \sigma^2_{u_m}$, and $F = E + (1 + 2\sqrt{3})\sigma^2_{u_m}$. Thus, for a large enough $n$, the combined estimator has lower variance than the backdoor estimator for all model parameter values.

\subsection{Comparison with the frontdoor estimator} \label{sec:appendix-combined-better-than-frontdoor}

In Section \ref{sec:combined-estimator-dominates-better-of-both}, we made the claim that
\begin{align*}
    \exists N, \text{s.t.}, \forall n > N, \Var(\widehat{a}_c \widehat{c}) \leq \Var(\widehat{a}_f \widehat{c}).
\end{align*}
In this case, by comparing Eqs. \ref{eq:frontdoor-finite-sample-variance} and \ref{eq:combined-finite-sample-upper-bound}, we have
\begin{align*}
    N = \frac{2 \left( \sigma^6_{u_m} + 2\sqrt{3}c^2 \sigma^4_{u_m} \sigma^2_{u_x} - c^4 \sigma^2_{u_m} \sigma^4_{u_x} + \left( D + \sigma^4_{u_m} \sqrt{ \sigma^4_{u_m} + 4\sqrt{3} c^2 \sigma^2_{u_m} \sigma^2_{u_x} - 2 c^4 \sigma^4_{u_x} } \right) \right)}{D^2},
\end{align*}
where $D = c^6 \sigma^6_{u_x}$. Thus, for a large enough $n$, the combined estimator has lower variance than the frontdoor estimator for all model parameter values.

\subsection{Combined estimator dominates the better of backdoor and frontdoor}
\label{sec:appedix-combined-estimator-dominates-better-of-both}

In this section, we provide more details for the claim in Section \ref{sec:combined-estimator-dominates-better-of-both} that the combined estimator can dominate the better of the backdoor and frontdoor estimators by an arbitrary amount. We show that the quantity
\begin{align*}
    R = \frac{\min \left\{ \Var(\widehat{ac})_\text{backdoor}, \Var(\widehat{a}_f\widehat{c}) \right\}}{\Var(\widehat{a}_c\widehat{c})}
\end{align*}
is unbounded.

We do this by considering the case when $\Var(\widehat{ac})_\text{backdoor} = \Var(\widehat{a}_f\widehat{c})$. Note that
\begin{align}
    & \Var(\widehat{ac})_\text{backdoor} = \Var(\widehat{a}_f\widehat{c}) \nonumber \\
    \implies & b = \sqrt{\frac{- D (-a^2 \sigma^4_{u_m} ( (n-2)d^2 \sigma^2_{u_w} + \sigma^2_{u_x} ) + (-(n-2)d^2 \sigma^2_{u_w} (\sigma^2_{u_m} - c^2 \sigma^2_{u_x}) + \sigma^2_{u_x} E ) \sigma^2_{u_y} ) }{\sigma^2_{u_w} \sigma^4_{u_x} (2 \sigma^2_{u_m} + (n-2)c^2 D )}}, \label{eq:appendix-b-equal-value}
\end{align}
where $D = d^2 \sigma^2_{u_w} + \sigma^2_{u_x}$, and $E = (n-2)c^2 \sigma^2_{u_x} - (n-4)\sigma^2_{u_m}$. Hence, if the parameter $b$ is set to the value given in Eq. \ref{eq:appendix-b-equal-value}, the backdoor and frontdoor estimators will have equal variance. We have to ensure that the value of $b$ is real. $b$ will be a real number if
\begin{align*}
    |c| &\leq \frac{\sigma_{u_m}}{\sigma_{u_x}} \sqrt{1 - \frac{2 \sigma^2_{u_x} }{ (n-2) D }}, \, \text{and} \\
    n &> 2.
\end{align*}
For the value of $b$ in Eq. \ref{eq:appendix-b-equal-value}, the quantity $R$ becomes
\begin{align}
    R &= \frac{\Var(\widehat{ac})_\text{backdoor}}{\Var(\widehat{a}_c\widehat{c})} \nonumber \\
        &\geq \frac{(n-2) D E (a^2 \sigma^2_{u_m} + \sigma^2_{u_y})}{ \sigma^2_{u_x} \left( (n-3) a^2 \sigma^2_{u_m} E + \sigma^2_{u_y} \left( \sigma^2_{u_m} + \sqrt{3} \sigma^2_{u_m} \left( r_1 r_2 + |c| (n-2) D \left( |c| + r_1 \frac{\sigma_{u_m}}{ \sqrt{(n-2) D} } \right) \right)  \right) \right) }, \nonumber
\end{align}
where $D = d^2 \sigma^2_{u_w} + \sigma^2_{u_x}$, $E = c^2 \sigma^2_{u_x} + \sigma^2_{u_m}$, $r_1 = \sqrt{\frac{n-3}{n-5}}$ and $r_2 = \sqrt{\frac{n-2}{n-4}}$.

$R$ does not depend on the parameter $b$. It is possible to set the other model parameters in a way that allows $R$ to take any positive value. In particular, it can be seen that as $\sigma_{u_x} \rightarrow 0$, $R \rightarrow \infty$, which shows that $R$ is unbounded.

\section{Combining Partially Observed Datasets}

\subsection{Cramer-Rao Lower Bound}
\label{sec:appendix-cramer-rao-bound}

We are interested in estimating the value of the product $ac$. 
Let $e = ac$. We reparameterize the likelihood in Eq. \ref{eq:partial_log_likelihood} 
by replacing $c$ with $e / a$. 
This simplifies the calculations and improves numerical stability. 
Now, we have the following eight unknown model parameters: $\{e,a,b,d,\sigma^2_{u_w},\sigma^2_{u_x},\sigma^2_{u_m},\sigma^2_{u_y}\}$.

In order to compute the variance of the estimate of parameter $e = ac$, 
we compute the Cramer-Rao variance lower bound. 
We first compute the Fisher information matrix (FIM) $\mathbf{I}$ 
for the eight model parameters:
\begin{align*}
    \mathbf{I} &= -E \begin{bmatrix}
    \frac{\partial^2 \mathcal{LL}}{\partial e^2} & \frac{\partial^2 \mathcal{LL}}{\partial e \partial a} & \frac{\partial^2 \mathcal{LL}}{\partial e \partial b} & \hdots & \frac{\partial^2 \mathcal{LL}}{\partial e \partial \sigma_{u_y}} \\
    \frac{\partial^2 \mathcal{LL}}{\partial a \partial e} & \frac{\partial^2 \mathcal{LL}}{\partial a^2} & \frac{\partial^2 \mathcal{LL}}{\partial a \partial b} & \hdots & \frac{\partial^2 \mathcal{LL}}{\partial a \partial \sigma_{u_y}} \\
    \vdots & \vdots & \vdots & \ddots & \vdots \\
    \frac{\partial^2 \mathcal{LL}}{\partial \sigma_{u_y} \partial e} & \frac{\partial^2 \mathcal{LL}}{\partial \sigma_{u_y} \partial a} & \frac{\partial^2 \mathcal{LL}}{\partial \sigma_{u_y} \partial b} & \hdots & \frac{\partial^2 \mathcal{LL}}{\partial^2 \sigma_{u_y}}
    \end{bmatrix}
\end{align*}
Let $\widehat{e}$ be the MLE. Since standard regularity conditions hold for our model (due to linearity and Gaussianity), the MLE is asymptotically normal. We can use the Cramer-Rao theorem to get the asymptotic variance of $\widehat{e}$. That is, for constant $k$, as $N \to \infty$, we have
\begin{align*}
    \sqrt{N}(\widehat{e} - e) &\overset{d}{\to} \mathcal{N}(0, V_e), \,\,\, \text{and} \\
    V_e &= (\mathbf{I}^{-1})_{1,1}.
\end{align*}

Below, we present the closed form expression for $V_e$.
Let $V_e = \frac{X}{Y}$. Then
\begin{align*}
    X = &(a^2 \sigma^2_{u_m}+\sigma^2_{u_y}) (-a^8 d^2 (k-1) \sigma^2_{u_w} (\sigma^2_{u_m})^5 (d^2 \sigma^2_{u_w}+\sigma^2_{u_x})^2 +a^6 (\sigma^2_{u_m})^3 (d^2 \sigma^2_{u_w}+\sigma^2_{u_x}) \\
    & (b^2 \sigma^2_{u_w} \sigma^2_{u_x} (c^2 d^4 (\sigma^2_{u_w})^2+d^2 \sigma^2_{u_w} (c^2 (k+1) \sigma^2_{u_x}+(-2 k^2+2 k+1) \sigma^2_{u_m})+\sigma^2_{u_x} (c^2 k \sigma^2_{u_x}+ \\
    & \sigma^2_{u_m})) + \sigma^2_{u_y} (d^2 \sigma^2_{u_w}+\sigma^2_{u_x}) (c^2 d^4 (\sigma^2_{u_w})^2+d^2 \sigma^2_{u_w} (c^2 (k+1) \sigma^2_{u_x}+(3-2 k) \sigma^2_{u_m})+ \\
    & k \sigma^2_{u_x} (c^2 \sigma^2_{u_x}+\sigma^2_{u_m})))-4 a^5 b c d (k-1) k \sigma^2_{u_w} \sigma^2_{u_x} (\sigma^2_{u_m})^3 (d^2 \sigma^2_{u_w}+\sigma^2_{u_x}) \\
    & (b^2 \sigma^2_{u_w} \sigma^2_{u_x}+\sigma^2_{u_y} (d^2 \sigma^2_{u_w}+\sigma^2_{u_x}))+a^4 (\sigma^2_{u_m})^2 (b^4 (\sigma^2_{u_w})^2 (\sigma^2_{u_x})^2 (c^2 d^4 (\sigma^2_{u_w})^2+ \\
    & d^2 \sigma^2_{u_w} (2 c^2 (-k^2+k+1) \sigma^2_{u_x}+(k+1) \sigma^2_{u_m})+\sigma^2_{u_x} (c^2 (-2 k^2+2 k+1) \sigma^2_{u_x}+2 \sigma^2_{u_m}))- \\
    & b^2 \sigma^2_{u_w} \sigma^2_{u_x} \sigma^2_{u_y} (d^2 \sigma^2_{u_w}+\sigma^2_{u_x}) (2 c^2 d^4 (k-2) (\sigma^2_{u_w})^2+d^2 \sigma^2_{u_w} (c^2 (4 k^2-3 k-5) \sigma^2_{u_x}+2 \\
    & (k^2-2 k-1) \sigma^2_{u_m})+\sigma^2_{u_x} (c^2 (4 k^2-5 k-1) \sigma^2_{u_x}-2 (k+1) \sigma^2_{u_m}))-(\sigma^2_{u_y})^2 (d^2 \sigma^2_{u_w}+\sigma^2_{u_x})^2 \\
    & (c^2 d^4 (2 k-3) (\sigma^2_{u_w})^2+d^2 \sigma^2_{u_w} (c^2 (2 k^2-k-3) \sigma^2_{u_x}+(k-3) \sigma^2_{u_m})+k \sigma^2_{u_x} (c^2 (2 k-3) \sigma^2_{u_x}- \\
    & 2 \sigma^2_{u_m})))-4 a^3 b c d (k-1) k \sigma^2_{u_w} \sigma^2_{u_x} (\sigma^2_{u_m})^2 \sigma^2_{u_y} (d^2 \sigma^2_{u_w}+\sigma^2_{u_x}) (b^2 \sigma^2_{u_w} \sigma^2_{u_x}+\sigma^2_{u_y} (d^2 \sigma^2_{u_w}+\sigma^2_{u_x}))+\\
    & a^2 \sigma^2_{u_m} (b^2 \sigma^2_{u_w} \sigma^2_{u_x}+\sigma^2_{u_y} (d^2 \sigma^2_{u_w}+\sigma^2_{u_x})) (b^4 (\sigma^2_{u_w})^2 (\sigma^2_{u_x})^2 (c^2 (d^2 \sigma^2_{u_w}-(k-2) \sigma^2_{u_x})+\sigma^2_{u_m})+ \\
    & b^2 \sigma^2_{u_w} \sigma^2_{u_x} \sigma^2_{u_y} (2 c^2 d^4 (\sigma^2_{u_w})^2+2 d^2 \sigma^2_{u_w} (c^2 (-k^2) \sigma^2_{u_x}+k (c^2 \sigma^2_{u_x}+\sigma^2_{u_m})+2 c^2 \sigma^2_{u_x})+\sigma^2_{u_x} (2 c^2 \\
    & (-k^2+k+1) \sigma^2_{u_x}+(k+1) \sigma^2_{u_m}))+(\sigma^2_{u_y})^2 (d^2 \sigma^2_{u_w}+\sigma^2_{u_x}) (d^2 \sigma^2_{u_w}+k \sigma^2_{u_x}) (\sigma^2_{u_m}-c^2 (2 k-3) \\
    & (d^2 \sigma^2_{u_w}+\sigma^2_{u_x})))+c^2 (b^2 \sigma^2_{u_w} \sigma^2_{u_x}+\sigma^2_{u_y} (d^2 \sigma^2_{u_w}+\sigma^2_{u_x}))^2 (b^4 (\sigma^2_{u_w})^2 (\sigma^2_{u_x})^2+b^2 \sigma^2_{u_w} \sigma^2_{u_x} \sigma^2_{u_y} \\
    & (2 d^2 k \sigma^2_{u_w}+k \sigma^2_{u_x}+\sigma^2_{u_x})+(\sigma^2_{u_y})^2 (d^2 \sigma^2_{u_w}+\sigma^2_{u_x}) (d^2 \sigma^2_{u_w}+k \sigma^2_{u_x}))),
\end{align*}
and
\begin{align*}
    Y = & (a^2 \sigma^2_{u_m} (d^2 \sigma^2_{u_w}+\sigma^2_{u_x})+b^2 \sigma^2_{u_w} \sigma^2_{u_x}+\sigma^2_{u_y} (d^2 \sigma^2_{u_w}+\sigma^2_{u_x})) (a^6 (-d^2) (k-1) \\
    & \sigma^2_{u_w} (\sigma^2_{u_m})^4 (d^2 \sigma^2_{u_w}+\sigma^2_{u_x})^2+a^4 (\sigma^2_{u_m})^2 (d^2 \sigma^2_{u_w}+\sigma^2_{u_x}) (b^2 \sigma^2_{u_w} \sigma^2_{u_x} (d^2 k \sigma^2_{u_w} \\
    & (c^2 \sigma^2_{u_x}-2 k \sigma^2_{u_m}+2 \sigma^2_{u_m})+\sigma^2_{u_x} (c^2 k \sigma^2_{u_x}+\sigma^2_{u_m}))+\sigma^2_{u_y} (d^2 \sigma^2_{u_w}+\sigma^2_{u_x}) (d^2 \sigma^2_{u_w} \\
    & (c^2 k \sigma^2_{u_x}-3 k \sigma^2_{u_m}+3 \sigma^2_{u_m})+k \sigma^2_{u_x} (c^2 \sigma^2_{u_x}+\sigma^2_{u_m})))-4 a^3 b c d (k-1) k \sigma^2_{u_w} \sigma^2_{u_x} (\sigma^2_{u_m})^2 \\
    & (d^2 \sigma^2_{u_w}+\sigma^2_{u_x}) (b^2 \sigma^2_{u_w} \sigma^2_{u_x}+\sigma^2_{u_y} (d^2 \sigma^2_{u_w}+\sigma^2_{u_x}))+a^2 \sigma^2_{u_m} (b^4 (\sigma^2_{u_w})^2 (\sigma^2_{u_x})^2 (\sigma^2_{u_x} \\
    & (\sigma^2_{u_m}-2 c^2 (k-1) k \sigma^2_{u_x})-d^2 (k-1) \sigma^2_{u_w} (2 c^2 k \sigma^2_{u_x}+\sigma^2_{u_m}))+2 b^2 \sigma^2_{u_w} \sigma^2_{u_x} \sigma^2_{u_y} (d^2 \sigma^2_{u_w}+\sigma^2_{u_x}) \\
    & (\sigma^2_{u_x} (\sigma^2_{u_m}-2 c^2 (k-1) k \sigma^2_{u_x})-2 d^2 (k-1) k \sigma^2_{u_w} (c^2 \sigma^2_{u_x}+\sigma^2_{u_m}))+(\sigma^2_{u_y})^2 (d^2 \sigma^2_{u_w}+\sigma^2_{u_x})^2 \\
    & (-d^2 (k-1) \sigma^2_{u_w} (2 c^2 k \sigma^2_{u_x}+3 \sigma^2_{u_m})-k \sigma^2_{u_x} (2 c^2 (k-1) \sigma^2_{u_x}+(k-2) \sigma^2_{u_m})))-4 a b c d (k-1) \\
    & k \sigma^2_{u_w} \sigma^2_{u_x} \sigma^2_{u_m} \sigma^2_{u_y} (d^2 \sigma^2_{u_w}+\sigma^2_{u_x}) (b^2 \sigma^2_{u_w} \sigma^2_{u_x}+\sigma^2_{u_y} (d^2 \sigma^2_{u_w}+\sigma^2_{u_x}))+b^6 c^2 k (\sigma^2_{u_w})^3 (\sigma^2_{u_x})^4+b^4\\
    & (\sigma^2_{u_w})^2 (\sigma^2_{u_x})^2 \sigma^2_{u_y} (d^2 \sigma^2_{u_w}+\sigma^2_{u_x}) (3 c^2 k \sigma^2_{u_x}-k \sigma^2_{u_m}+\sigma^2_{u_m})+b^2 \sigma^2_{u_w} \sigma^2_{u_x} (\sigma^2_{u_y})^2 (d^2 \sigma^2_{u_w}+\sigma^2_{u_x}) \\
    & (d^2 k \sigma^2_{u_w}  (3 c^2 \sigma^2_{u_x}-2 (k-1) \sigma^2_{u_m})+\sigma^2_{u_x} (3 c^2 k \sigma^2_{u_x}-k^2 \sigma^2_{u_m}+\sigma^2_{u_m}))+(\sigma^2_{u_y})^3 (d^2 \sigma^2_{u_w}+\sigma^2_{u_x})^2 \\
    & (d^2 \sigma^2_{u_w} (c^2 k \sigma^2_{u_x}-k \sigma^2_{u_m}+\sigma^2_{u_m})+k \sigma^2_{u_x} (c^2 \sigma^2_{u_x}-k \sigma^2_{u_m}+\sigma^2_{u_m}))),
\end{align*}
where $k = \frac{P}{N}$.

\subsection{Comparison with frontdoor and backdoor estimators}
\label{sec:appendix-cramer-rao-comparison}

\begin{figure}[t]
\begin{center}
\begin{subfigure}{0.32\columnwidth}
\includegraphics[width=0.9\columnwidth]{figures/cramer_rao_middle_0.303.eps}
\caption{}
\label{fig:appendix-cramer-rao-a}
\end{subfigure}
\begin{subfigure}{0.32\columnwidth}
\includegraphics[width=0.9\columnwidth]{figures/cramer_rao_middle_0.505.eps}
\caption{}
\label{fig:appendix-cramer-rao-b}
\end{subfigure}
\begin{subfigure}{0.32\columnwidth}
\includegraphics[width=0.9\columnwidth]{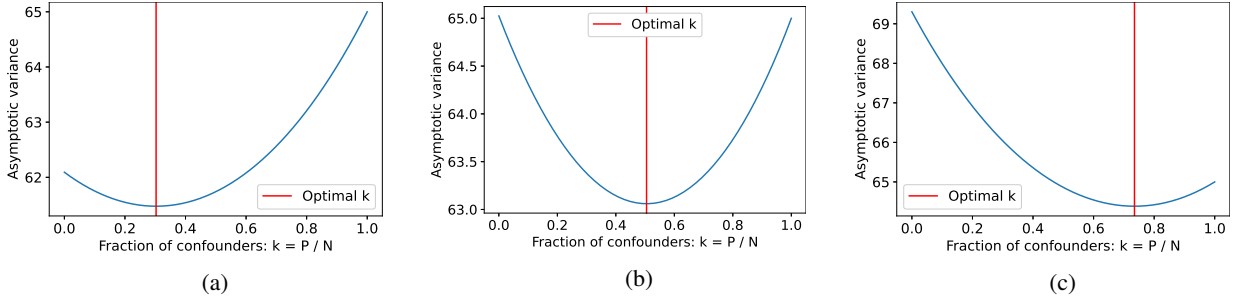}
\caption{}
\label{fig:appendix-cramer-rao-c}
\end{subfigure}
\caption{Cases where collecting a mix of confounders and mediators is better than collecting only confounders or mediators.}
\label{fig:appendix-cramer-rao}
\end{center}
\end{figure}

In this section, we show some examples of regimes where the combining partially observed datasets results in lower variance than applying either of the backdoor or frontdoor estimator even when the total number of samples are the same. In other words, there exist settings of model parameters such that, for some $k \in (0, 1)$, we have
\begin{align*}
    V_e \leq \Var(\sqrt{N} \widehat{ac})_{\text{backdoor}},\,\, \text{and} \,\,
    V_e \leq \Var(\sqrt{N} \widehat{a}_f \widehat{c}).
\end{align*}

Figure \ref{fig:appendix-cramer-rao} shows three examples where the optimal value of $k$ is between $0$ and $1$. We plot the variance as predicted by the expression for $V_e$ versus the value of $k$. The plots show that in some cases, it is better to collect a mix of confounders and mediators rather than only mediators or only confounders. The expression for $V_e$ in the previous section allows us to verify that. This happens when the variance of the frontdoor and backdoor estimators do not differ by too much. 

In Figure \ref{fig:appendix-cramer-rao-a}, the model parameters are $\{a=10, b=3.7, c=5, d=5, \sigma^2_{u_w}=1, \sigma^2_{u_x}=1, \sigma^2_{u_m}=0.64, \sigma^2_{u_y}=1 \}$. In this case, the variance of the frontdoor estimator is lower than the backdoor estimator. Despite this, it is not optimal to only collect mediators. The optimal value of $k$ is $0.303$, that is, $30\%$ of the collected samples should be confounders and the rest should be mediators to achieve lowest variance.

In Figure \ref{fig:appendix-cramer-rao-b}, the model parameters are $\{a=10, b=3.955, c=5, d=5, \sigma^2_{u_w}=1, \sigma^2_{u_x}=1, \sigma^2_{u_m}=0.64, \sigma^2_{u_y}=1 \}$. In this case, the variance of the frontdoor estimator is almost equal to that of the backdoor estimator. The optimal ratio $k$ is $0.505$, that is, we should collect the same of amount of confounders as mediators.

In Figure \ref{fig:appendix-cramer-rao-c}, the model parameters are $\{a=10, b=4.3, c=5, d=5, \sigma^2_{u_w}=1, \sigma^2_{u_x}=1, \sigma^2_{u_m}=0.64, \sigma^2_{u_y}=1 \}$. In this case, the variance of the frontdoor estimator is greater than the backdoor estimator. The optimal ratio $k$ is $0.735$, that is, we should collect the more confounders than mediators.

\subsection{Parameter initialization for finding the MLE}
\label{sec:appendix-cramer-rao-param-init}

The likelihood in Eq. \ref{eq:partial_log_likelihood} is non-convex. 
As a result, we cannot start with arbitrary initial values 
for model parameters because we might encounter a local minimum. 
To avoid this, we use the two datasets to initialize our parameter estimates. 
Each of the eight parameters can be identified 
using only data from one of the datasets.
For example, $d$ can be initialized using 
the revealed-confounder dataset 
(via OLS regression of $X$ on $W$). 
The parameter $e$ is can be identified using either dataset, 
so we pick the value with lower bootstrapped variance.

After initializing the eight model parameters, 
we run the Broyden–Fletcher–Goldfarb–Shanno (BFGS) algorithm \citep{fletcher2013practical}
to find model parameters that minimize the negative log-likelihood.

\section{More details on experiments}\label{sec:appendix-synthetic-experiments}

Here we provide more details for how results in Table \ref{table:synthetic-theoretical-vs-empirical-variance} are generated.
We initialize the model parameters 
by sampling $200$ times from the following distributions:
\begin{align}\label{eq:synthetic_data_param_distributions}
\begin{split}
    a,b,c,d &\sim \text{Unif}[-10, 10]\\
    \sigma^2_{u_w},\sigma^2_{u_x},\sigma^2_{u_m},\sigma^2_{u_y} &\sim \text{Unif}[0.01, 2].
\end{split}
\end{align}
For each initialization, we compute the Mean Absolute Percentage Error (MAPE) of the theoretical variance as a predictor of empirical variance:
\begin{align*}
    \text{MAPE} = \frac{\left| \Var_{\text{theoretical}} - \Var_{\text{empirical}} \right|}{\Var_{\text{empirical}}} * 100\%
\end{align*}
We report the mean and standard deviation of the MAPE across $1000$ realizations of datasets
sampled from Eq. \ref{eq:synthetic_data_param_distributions}.  We find that the theoretical variance is close 
to the empirical variance even for small sample sizes
(Table \ref{table:synthetic-theoretical-vs-empirical-variance}).